\documentclass[fleqn,usenatbib]{mnras}
\usepackage{newtxtext,newtxmath}
\usepackage[T1]{fontenc}
\usepackage{ae,aecompl}
\pdfoutput=1
\usepackage{multirow}
\usepackage{graphicx}	
\usepackage{amsmath}	
\usepackage{amssymb}	
\usepackage{etoolbox}

\title[SF-Gas relations in nearby spirals]{A panchromatic spatially resolved analysis of nearby galaxies - II. The main sequence - gas relation at sub-kpc scale in grand-design spirals}

\author[Morselli L. et al.]{
L. Morselli$^{1,2}$\thanks{E-mail: laura.morselli@unipd.it},
G. Rodighiero$^{1,2}$,
A. Enia$^{1,2}$,
E. Corbelli$^{3}$,
V. Casasola$^{4,3}$, 
\newauthor
L. Rodr\'iguez-Mu{\~n}oz$^{1}$,
A. Renzini$^{2}$,
S. Tacchella$^{5}$,
I. Baronchelli$^{3}$,
S. Bianchi$^{3}$,
\newauthor
P. Cassata$^{1,2}$,
A. Franceschini$^{1}$,
C. Mancini$^{1,2}$,
M. Negrello$^{6}$,
P. Popesso$^{7}$,
M. Romano$^{1,2}$
\\
\\
$^{1}$ Dipartimento di Fisica e Astronomia, Universit{\`a} di Padova, vicolo dell'Osservatorio 3, I-35122 Padova, Italy\\
$^{2}$ INAF $-$ Osservatorio Astrofisico di Padova, vicolo dell'Osservatorio 5, I-35122 Padova, Italy\\
$^{3}$ INAF $-$ Osservatorio Astrofisico di Arcetri, Largo E. Fermi 5, I-50125, Firenze, Italy\\
$^{4}$ INAF $-$ Istituto di Radioastronomia, Via P. Gobetti 101, I-40129, Bologna, Italy\\
$^{5}$ Center for Astrophysics | Harvard \& Smithsonian, 60 Garden St, Cambridge, MA 02138, USA\\
$^{6}$ School of Physics and Astronomy, Cardiff University, The Parade, Cardiff CF24 3AA, UK\\
$^{7}$ Excellence Cluster Universe, Boltzmann strasse 2, D-85748 Garching bei M\"unchen, Germany
}

\date{Accepted 2020 June 18. Received YYY; in original form ZZZ}

\pubyear{2020}

\begin{document}
\label{firstpage}
\pagerange{\pageref{firstpage}--\pageref{lastpage}}
\maketitle

\begin{abstract}

In this work we analyse the connection between gas availability and the position of a region with respect to the spatially resolved main sequence (MS) relation. Following the procedure presented in \citet{2020MNRAS.tmp..429E}, for a sample of five face-on, grand design spiral galaxies located on the MS we obtain estimates of stellar mass and star formation rate surface densities ($\Sigma_{\star}$ and $\Sigma_{\rm{SFR}}$) within cells of 500pc size. Thanks to HI 21cm and $^{12}$CO(2-1) maps of comparable resolution, within the same cells we estimate the surface densities of the atomic ($\Sigma_{\rm{HI}}$) and molecular ($\Sigma_{\rm{H_2}}$) gas and explore the correlations among all these quantities. $\Sigma_{\star}$, $\Sigma_{\rm{SFR}}$ and $\Sigma_{\rm{H_2}}$ define a 3D relation whose projections are the spatially resolved MS, the Kennicutt-Schmidt law and the molecular gas MS. We find that $\Sigma_{\rm{H_2}}$ steadily increases along the MS relation and is almost constant perpendicular to it. $\Sigma_{\rm{HI}}$ is nearly constant along the MS and increases in its upper envelope. As a result, $\Sigma_{\rm{SFR}}$ can be expressed as a function of $\Sigma_{\star}$ and $\Sigma_{\rm{HI}}$, following the relation $\log\Sigma_{\rm{SFR}}$ = 0.97$\log\Sigma_{\star}$ + 1.99$\log\Sigma_{\rm{HI}}$ - 11.11. We show that the total gas fraction significantly increases towards the starburst regions, accompanied by a weak increase in star formation efficiency. Finally, we find that H$_2$/HI varies strongly with the distance from the MS, dropping dramatically in regions of intense star formation, where the UV radiation from newly formed stars dissociates the H$_2$ molecule, illustrating the self-regulating nature of the star formation process.

\end{abstract}




\section{Introduction}

In the current model of galaxy formation and evolution stars form in dense clouds of molecular gas, thanks to the interplay of different physical mechanisms (magnetic fields, turbulence, shielding, feedback). Despite its complexity, this interplay translates in tight correlations between different physical quantities: \textit{i)} between the surface density of the star formation rate ($\Sigma_{\rm SFR}$) and the surface density of the gas ($\Sigma_{\rm gas}$), and \textit{ii)} between the stellar mass surface density ($\Sigma_{\star}$) and $\Sigma_{\rm SFR}$. The first relation, originally formulated by \citet{1959ApJ...129..243S} using the gas volume density and the number of stars formed in the solar neighborhood, was subsequently derived by \citet{1998ARA&A..36..189K} for radially averaged surface densities in external galaxies, and it is thus called the Kennicutt-Schmidt (KS) relation. The second, called main sequence (MS), was initially found using integrated quantities of star-forming galaxies (thus the total SFR and stellar mass M$_{\star}$) in \citet{2004MNRAS.351.1151B} for local galaxies and later confirmed for high-redshift galaxies by several works \citep[e.g.][]{2007ApJS..173..267S, 2007ApJ...660L..47N,2007A&A...468...33E,2007ApJ...670..156D}. As both relations are intrinsically related to the process of star formation and thus to galaxy evolution as a whole, and because they are fundamental ingredients of theoretical models and simulations, they have been intensively studied in the past \citep[e.g.][]{2000ApJ...536..173T,2003MNRAS.346.1215B,2003MNRAS.339..312S,2005ApJ...630..250K,2011ApJ...739L..40R,2012ApJ...745...69K,2012ApJ...754L..29W, 2012ARA&A..50..531K,2013ApJ...777L...8K,2014ApJS..214...15S,2014MNRAS.445..581H,2015ApJ...800...20G,2015A&A...575A..74S,2016ApJ...820L...1K,2017ApJ...847...76S,2016MNRAS.457.2790T,2018MNRAS.478.3653O,2018A&A...615A.146P,2019MNRAS.483.3213P,2019A&A...626A..61M}.

The KS law relates the fuel of star formation to its end product, stars; its shape has important effects on the depletion time of the gas (t$_{\rm depl}$ = M$_{\rm gas}$/SFR, with M$_{\rm gas}$ the total gas mass), or equivalently on the efficiency of the star formation process (SFE = t$_{\rm depl}\ ^{-1}$ = SFR/M$_{\rm gas}$). In one of the earliest works, \citet{1998ARA&A..36..189K} finds a super linear correlation (slope = 1.4-1.5) between the total gas and the SFR surface densities. Following this result, several papers investigate the relation between star formation and gas availability, considering different gas phases and star formation tracers, as well as exploring this link at different cosmic epochs \citep[e.g.][]{2009ApJ...696.1834W, 2010MNRAS.407.2091G, 2010Natur.463..781T, 2012ApJ...746...69G}. \citet{2008AJ....136.2846B}, \citet{2008AJ....136.2782L, 2013AJ....146...19L} and \citet{2011AJ....142...37S} exploit molecular and neutral gas observations of nearby galaxies to investigate how the relation between gas and star formation activity varies within galaxies and as a function of local and integrated properties. Overall, their findings indicate that the connection between star formation and molecular gas is a linear relation (i.e., slope $\sim$ 1), thus implying a constant molecular SFE and t$_{\rm depl}$ (around 2.2 Gyr). \citet{2013AJ....146...19L} find second order variations in the molecular gas t$_{\rm depl}$ and study how some of them can be related to variations in the $\alpha_{\rm CO}$ conversion factor between CO luminosity and H$_2$ mass, while further variability might arise as a consequence of galaxy properties. \citet{2010AJ....140.1194B}, instead, study the relation between recent SF activity and HI outside the optical disc, in regions where HI represents the totality of the ISM, and find significant spatial correlation between FUV (tracing recent dust-unobscured star formation) and HI density. They also find that the SFE (t$_{\rm depl}$) decreases (increases) with increasing radius. Similarly, \citet{2015MNRAS.449.3700R} study the spatially resolved KS relation on sub-kpc and kpc scales in the HI dominated regions of nearby spirals and irregular galaxies and find that gas consumption time-scales are longer compared to H$_2$ dominated regions (lower SFE). Other works investigated, at earlier cosmic epoch, the spatially resolved \citep[e.g.][]{2013ApJ...773...68G, 2013A&A...553A.130F} and integrated \citep[][e.g.]{2019A&A...622A.105F} KS relation. In particular, \cite{2019A&A...622A.105F} obtain a linear galaxy-averaged molecular KS relation, implying that galaxies at different cosmic epochs have similar star formation timescales. This is consistent with the results of \citet{2020MNRAS.491L..51P}, that find that the sSFR (SFR/M$_{\star}$) and M$_{\rm H_2}$/M$_{\star}$ have the same redshift evolution, thus implying a linear KS law. On the other hand, galaxy-to-galaxy variations in the molecular gas - star formation relation have also been reported. For example, \citet{2013ApJ...769...55F} and \citet{2013MNRAS.430..288S, 2014MNRAS.442.2208S, 2014MNRAS.437L..61S} find evidence for a sub-linear relation within galaxies and for the combined samples. Also \citet{2015A&A...577A.135C} find galaxy-to-galaxy variations of the spatially resolved KS relation, and underline that the slope can be both sub-linear and super-linear, depending on the spatial scale.  \cite{2019ApJ...872...16D} revisit the integrated KS law in local, normal star-forming galaxies and find that spirals lie on a tight log-linear relation with slope 1.41$\pm$0.07 (when considering both the neutral and molecular gas) while dwarfs populate the region below it.

\begin{table*}
	\centering
	\caption{Properties of the galaxies in our sample: (1) galaxy name; (2,3) RA and Dec coordinates in J2000; (4) total M$_{\star}$ from SED fitting; (5) total SFR computed from Eq.\,\ref{eq:sfr} with L$_{\rm UV}$ and L$_{\rm IR}$ from SED fitting; (6,7) distance in Mpc and radius in kpc at which the optical surface brightness falls below 25 mag arcsec$^{-2}$, both taken from the HyperLEDA data base; (8) total HI mass in R$_{25}$; (9) total H$_2$ mass in R$_{25}$; 10) morphological T type, taken from HyperLEDA; (11) central metallicity computed using the O3N2 index, and taken from the DustPedia archive. }
	\label{tab:sample}
	\begin{tabular}{lccccccccccc}
		\hline
		Galaxy Name     & RA & DEC & logM$_{\star}$ & SFR & D & R$_{25}$ & logM$_{\rm HI}$ & logM$_{\rm H_2}$ & T & 12 + log(O/H) \\
		                & [deg] & [deg] & [M$_{\odot}$] & [M$_{\odot}$/yr]   & [Mpc] & [kpc]   & [M$_{\odot}$] & [M$_{\odot}$]   \\
		\hline
        NGC0628 (M74) & 24.174 & 15.7833 & 10.31$\pm$0.15 & 1.78$\pm$0.41 & 10.14 & 14.74 & 9.54$\pm$0.18 & 9.39$\pm$0.20 & 5.2 & 8.693$\pm$0.001\\ 
        NGC3184 & 154.5708 & 41.4244 & 10.13$\pm$0.10 & 1.02$\pm$0.10 & 11.64 & 12.55 &  9.38$\pm$0.14 & 9.20$\pm$0.24 & 5.9 & 8.766$^{+0.014}_{-0.013}$\\
        NGC5194 (M51a) & 202.8025 & 47.1952 & 10.74$\pm$0.20 & 3.68$\pm$0.26 & 8.59 & 17.23  & 9.43$\pm$0.17 & 9.86$\pm$0.22 & 4.0 & 8.824$^{+0.017}_{-0.016}$\\
        NGC5457 (M101) & 210.8025 & 54.3491 & 10.37$\pm$0.13 & 3.00$\pm$0.15 & 7.11 & 24.81  & 10.15$\pm$0.14 & 9.41$\pm$0.18 & 5.9 & 8.528$^{+0.006}_{-0.007}$\\
        NGC6946 & 308.71905 & 60.15361 & 10.61$\pm$0.13 & 3.51$\pm$0.15 & 6.73 & 24.81 & 9.41$\pm$0.15 & 9.74$\pm$0.17 & 5.9 & 8.746$^{+0.067}_{-0.070}$ \\
		\hline
	\end{tabular}
\end{table*}

The second fundamental relation, the MS, relates stars that have already formed to the ongoing SFR. The existence of the MS up to z$\sim$4 \citep[characterised by a non-evolving slope and scatter and an increasing normalisation with increasing redshift, e.g.][]{2014ApJS..214...15S,2019MNRAS.483.3213P}
 was interpreted in the framework of gas-regulated galaxy evolution, according to which galaxies grow along the MS thanks to the continuous replenishment of their gas supply \citep[e.g.][]{2009Natur.457..451D,2010ApJ...718.1001B,2013ApJ...772..119L}. The observation of outliers located above the MS relation \citep[starburst are generally classified in literature as galaxies having a SFR that is a factor 4 higher that the MS value at fixed stellar mass, e.g.][]{ 2011ApJ...739L..40R} at different cosmic epochs sparkled the interest on whether these sources: {\it i)} have larger gas reservoirs \citep[thus a higher gas fraction, $f_{\rm gas}$ = M$_{\rm gas}$/(M$_{\rm gas}$+M$_{\star})$, e.g.][]{2017MNRAS.471.2124L}, {\it ii)} are more efficient in converting gas into stars \citep[thus a higher SFE, e.g.][]{1997ApJ...478..144S, 2010ApJ...714L.118D,2015ApJ...812L..23S,2018ApJ...867...92S,2020MNRAS.492.6027E}, or {\it iii)} a combination of both \citep[e.g.][]{2011MNRAS.415...32S,2012ApJ...758...73S,2014MNRAS.443.1329H,2017ApJ...837..150S, 2018ApJ...853..179T}. In the recent years, the advent of large integral field spectroscopic (IFS) surveys revealed that the integrated MS relation originates at smaller scales (up to the sizes of molecular clouds), thus implying that the star formation process is regulated by physical process that act on sub-galactic scales \citep{2016ApJ...821L..26C,2017ApJ...851L..24H,2017ApJ...851...18L,2017MNRAS.469.2806A,2018MNRAS.475.5194M,2018ApJ...865..154H,2019MNRAS.488.3929C,2019MNRAS.488.1597V,2020MNRAS.492...96B,2020MNRAS.tmp..429E}. Despite a general consensus on the existence of the spatially resolved MS, the slope, intercept and scatter of the relation vary significantly among different works, depending on the sample selection, SFR indicator, dust correction, and fitting procedure. Moreover, some authors find that the spatially resolved relation vary dramatically from galaxy to galaxy. Recently, the combination of MaNGA \citep[Mapping Nearby Galaxies at APO,][]{2015ApJ...798....7B} and ALMaQUEST (the ALMA-MaNGA QUEnching and STar formation survey) allowed the study of the link between the spatially resolved MS and gas reservoirs and the investigation of the nature of starburst regions within galaxies. By analysing 14 MS galaxies, \citet{2019ApJ...884L..33L} suggest that the MS relation originates from two more fundamental relations: the molecular KS and the so-called \textit{molecular gas main sequence} (MGMS), a relation between $\Sigma_{\star}$ and $\Sigma_{\rm H2}$. \citet{2020MNRAS.493L..39E} exploit 34 galaxies in ALMaQUEST to study the nature of variations in the SFR on kpc scales. They find that while the average SFR is regulated by the availability of molecular gas, the scatter of the spatially resolved MS (and thus variations with respect to the average SFR value) originates in variations of the SFE. \citet{2019MNRAS.488.1926D}, using optical IFU and CO observations collected in the EDGE-CALIFA survey \citep[CARMA Extragalactic Database for Galaxy Evolution][see also \citealt{2020MNRAS.492.2651B}]{2017ApJ...846..159B}, find that $\Sigma_{\rm SFR}$ is a function of both $\Sigma_{\star}$ and $\Sigma_{\rm H_2}$ but, differently from \citet{2019ApJ...884L..33L}, the relation with the stellar mass is statistically more significant than the one with the molecular gas. Early works on the formation of molecular hydrogen in the ISM, such as \citet{1993ApJ...411..170E} and \citet{2006ApJ...650..933B}, have underlined the role of the disc hydrostatic pressure, and hence of $\Sigma_{\star}$, in promoting the formation of molecules. Also the works of \citet{2011ApJ...733...87S} and \citet{2018ApJ...853..149S}, that propose an {\it extended KS law} expressed as a proportionality between $\Sigma_{SFE}$ and $\Sigma_{\star}$, emphasise the role of existing stars in setting the current production of stars, which indeed is the very nature of the MS.

In this paper, we build on the work presented in \citet[][hereafter Paper I]{2020MNRAS.tmp..429E} and analyse the sub-kpc relation between the surface densities of star formation, gas in different phases, and stellar mass in 5 local grand-design spirals. In Paper I we exploit multiwavelength observations in more than 20 photometric bands to obtain spatially resolved estimates of $\Sigma_{\star}$ and $\Sigma_{\rm{SFR}}$ on different physical scales, from few hundred parsecs to 1.5 kpc, via SED fitting. We use these estimates to study the spatially resolved MS relation and find the slope to be consistent for different spatial scales, as well as with the slope of the integrated relation. Here, we aim at analysing under which gas properties different spatial regions populate different loci of the spatially resolved MS, thus trying to understand whether the SFR is more connected to the gravity of the disc (dominated by stars up to $\sim$ 2/3 of the optical radius) or with the availability of fuel, or a combination of both. We exploit observations in more than 20 photometric bands to derive accurate SFR and M$_{\star}$ maps to compare to HI and H$_2$ maps. We discuss the origin of the spatially resolved MS, in terms of its slope and scatter. 

The structure of the paper is the following: in Section\,\ref{sec:Data} we give a short description of the data used in this work; in Section\,\ref{sec:Results} we present our results at 500 pc resolution; in Section \ref{sec:Discussion} we analyse the implications of our results on the existence of the MS relation and on how SFE and $f_{gas}$ vary with varying SFR. Finally, in Section \ref{sec:Conclusion} we summarise our findings. The assumed IMF is \citet{2003PASP..115..763C}, cosmology is $\Lambda$CDM with parameters from \citet{2016A&A...594A..13P}.

\begin{figure*}
	\includegraphics[width=0.99\textwidth]{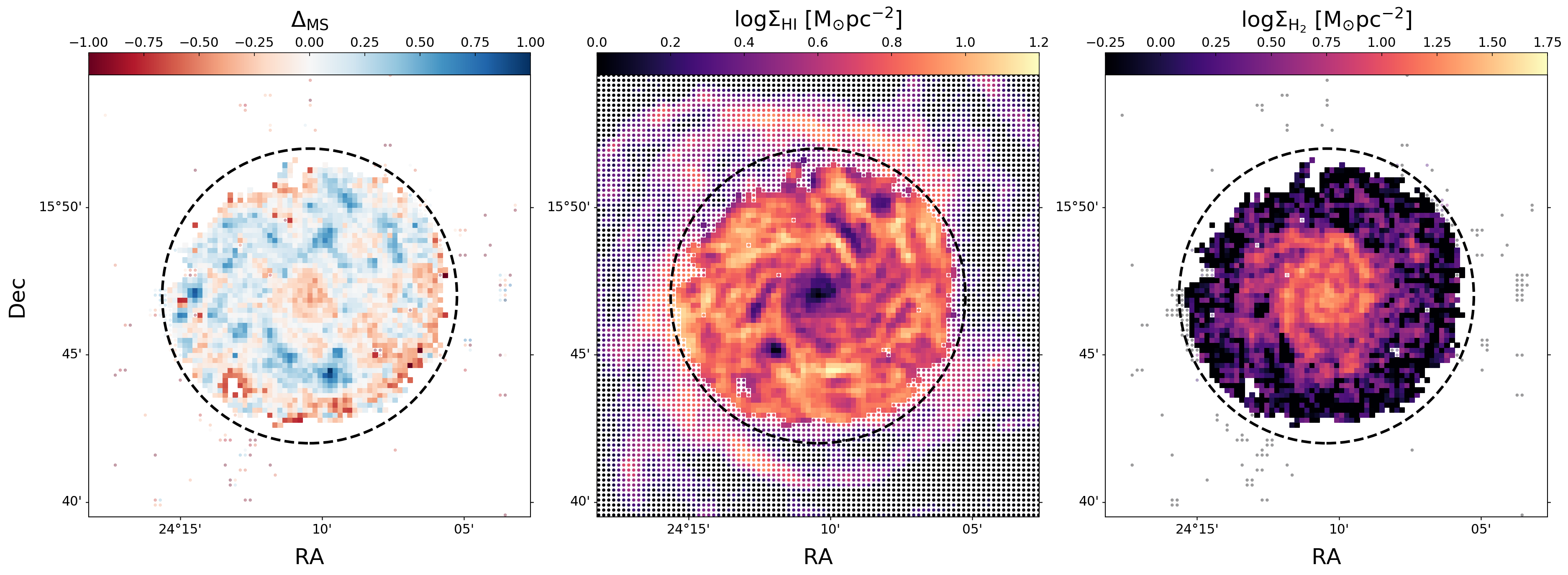}
   \caption{Spatially resolved properties of NGC0628: $\Delta_{\mathrm{MS}}$ (left panel), $\log\Sigma_{\mathrm{HI}}$ (central panel), and  $\log\Sigma_{\mathrm{H_2}}$ (right panel). The dashed circle marks the optical radius of the galaxy.} 
    \label{fig:0628}
\end{figure*} 

\section{Data}\label{sec:Data}
This work is based on multiwavelength observations of five nearby face-on spiral galaxies: NGC0628, NGC3184, NGC5194, NGC5457, and NGC6946. Four out of five galaxies are in common with Paper I: they are the ones observed in 23 photometric bands, and included in the THINGS and HERACLES surveys. NGC6946 was initially excluded from the Paper I sample since it lacked optical observations (the five Sloan optical filters). We tested for the sample in Paper I how the SED fitting routine results change excluding these five photometric points, finding that they are nearly unchanged. Following this, we are including NGC6946 in this analysis. The observations in 23 different bands (18 for NGC6946) have been collected in the DustPedia \footnote{The DustPedia data base is available at http://DustPedia.astro.noa.gr} \citep{2017PASP..129d4102D,2018A&A...609A..37C} archive; more details on the data set can be found in Paper I and references therein. The main properties of the galaxies in this sample are shown in Table\,\ref{tab:sample}. We highlight two properties of our sample: {\it i)} according to the integrated SFR and M$_{\star}$ values, the objects in our sample are MS galaxies, located within 0.2 dex from the relation obtained in Paper I, and {\it ii)} three out of five sources are classified as SAB spirals (NGC3184, NGC5457 and NGC6946), thus show evidence of a bar component.

\subsection{SFRs, stellar mass, and distance from the MS}
\label{2e1}
The spatially resolved measurements at 500 pc resolution of $\Sigma_{\star}$, $\Sigma_{\rm{SFR}}$ and distance from the MS ($\Delta_{\rm MS}$), have been obtained following the procedure presented in Paper I. Briefly, we select 8 nearby, face-on, grand design spiral MS galaxies with $\log$M$_{\star} \sim$ 10.4-10.6 M$_{\odot}$, and perform spatially resolved SED fitting to 23 photometric bands using \textsc{magphys} \citep{2008MNRAS.388.1595D}. In particular, we performed SED fitting on cells having two different side measurements: 8 arcsec (thus a varying physical scale between 290pc and 700pc, depending on the distance of the galaxies) and 1.5 kpc. Here we implement an improved procedure, and performed SED fitting at a common resolution of 500 pc (as discussed in Paper I, these scales are higher than the ones where the energy-balance criterion holds, $\sim 200-400$ pc). The procedural improvements are the following: \textit{i)} we estimate the noise on the photometry of each cell from the rms maps (while in Paper I we used the DustPedia photometry signal-to-noise ratio); \textit{ii)} if a cell has more than 10 bands with SNR $< 2$ it is automatically excluded from the SED fitting procedure, thus reducing computational time. These improvements influence the $\chi^2$ estimation in \textsc{magphys}, increasing the number of accepted points within the optical radius, and leading to cleaner results in the outer part of galaxies, where the photometry is fainter. The slope and intercept of the spatially resolved MS given in Paper I do not change when these improvements are implemented in the pipeline. 

In each cell, the SFR is computed as the sum of unobscured (SFR$_{\rm UV}$) and obscured (SFR$_{\rm IR}$) star formation activity, obtained using the relations of \citet{2001ApJ...548..681B} and \citet{1998ARA&A..36..189K} (reported to  Chabrier IMF): 
\begin{equation}
    {\rm SFR} = 0.88\cdot 10^{-28}L_{\rm{UV}} + 2.64\cdot 10^{-44}L_{\rm IR},
    \label{eq:sfr}
\end{equation}
where $L_{\rm{UV}}$ and $L_{\rm IR}$ are taken from the best fitting SED and are the luminosity (in erg s$^{-1}$ Hz$^{-1}$) at 150nm and the one (in erg s$^{-1}$) integrated between 8 and 1000$\mu$m, respectively. As shown in Fig.\,3 of Paper I, the SFR computed following Eq.\ref{eq:sfr} and the one that \textsc{magphys} gives as output, are highly consistent. Here, for consistency with Paper I, we use the SFRs estimated with Eq.\ref{eq:sfr}, but the results would not change when considering the SFRs given as output of the SED fitting procedure. 

As the sample of galaxies used here differs from the one of Paper I, we decided to recompute the spatially resolved MS for this sample, but following the same procedure, i.e., 1) by fitting with a log-linear relation the median values of log$\Sigma_{\rm SFR}$ in bins of log$\Sigma_{\star}$, using {\sc emcee} \citep{2013PASP..125..306F} and considering 10 bins in the log$\Sigma_{\star}$ range [6.5:8.5] M$_{\odot}$pc$^{-2}$, plus an additional bin to include the few points between [8.5:9.5]M$_{\odot}$pc$^{-2}$, and 2) by implementing an orthogonal distance regression (ODR) technique. The slope and intercept of the MS are 0.76($\pm 0.20$) and -8.15($\pm 1.63$) with the first method, and 0.87($\pm 0.01$) and -8.94($\pm 0.06$) with the second. These estimates are consistent with the ones in Paper I. In the following analysis we make use of the distance from the MS relation computed with the binning technique, but our results do not change when considering the ODR MS relation. 

For each region we compute the distance from the MS as the difference between $\log\Sigma_{\rm SFR}$ and the MS value (in log) estimated for the $\Sigma_{\star}$ of the region, thus: $\Delta_{\rm MS}$ = $\log\Sigma_{\rm SFR}$ - (0.76$\log\Sigma_{\star}$ - 8.15). The left panel of Fig.\,\ref{fig:0628} shows, as an example, the $\Delta_{\rm MS}$ map of NGC0628; cells in red are located below the spatially resolved MS, while the ones in blue are located above the relation  (the $\Delta_{\rm MS}$ maps of the other galaxies in the sample are shown in Appendix \ref{sec:a0}). Within the optical radius (the dashed circle) we are able to recover most of the cells, especially at $r<0.9$ R$_{25}$ \footnote{R$_{25}$ is defined as the length of the projected semi-major axis of a galaxy at the isophotal level 25 mag/arcsec$^2$ in the B-band and it is taken from the HyperLEDA data base \citep{2014A&A...570A..13M}.}. We emphasise here that the MS relation we obtain is well defined also in the outer parts of the optical disc, where the SFR and M$_{\star}$ are small in absolute values. We refer the reader to Paper I for details on the SED fitting procedure, as well as for how the spatially resolved MS relation is obtained. 

\begin{figure*}
	\includegraphics[width=0.99\textwidth]{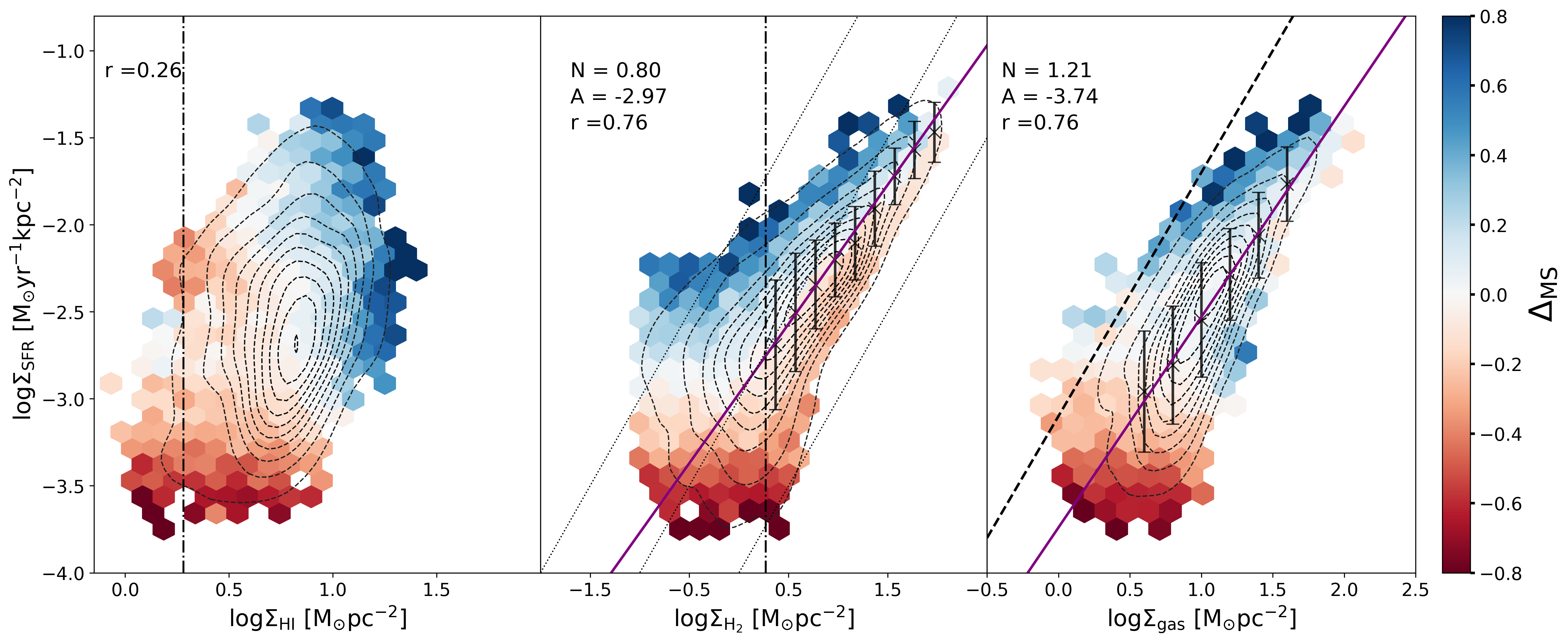}
    \caption{Relations between $\Sigma_{\rm SFR}$ and $\Sigma_{\rm HI}$ (left panel), $\Sigma_{\rm H_2}$ (middle panel) and  $\Sigma_{\rm gas}$ (for the total HI+H$_2$ gas, right panel) colour coded as a function of the median value of $\Delta_{\rm MS}$ in each bin. Only bins containing a minimum number of 3 cells are shown in the plot. The dashed contours encircle the areas of the plane containing from 10\% to 90\% of the data, at steps of 10\%. The sensitivity limits are represented by the dotted-dashed black lines. The purple solid lines in the middle and right panel represent the best fit to the data obtained fitting the points marked with crosses; the corresponding slope (N) and intercept (A) are written in the panels, together with the Spearman correlation coefficient \rm{r}. In the central panel, the dotted lines mark constant molecular t$_{\rm depl}$ of 10$^{8}$, 10$^{9}$, and 10$^{10}$ yr from top to bottom. In the right panel, the dashed black line is the fit to local ULIRGs and SMGs taken from \citet{2010ApJ...714L.118D}.}
    \label{fig:bigiel}
\end{figure*}

\subsection{Neutral gas: HI 21cm observations}
Neutral hydrogen mass surface densities ($\Sigma_{\rm{HI}}$) are measured from 21cm maps available from the THINGS survey \citep[The HI Nearby Galaxy Survey,][]{2008AJ....136.2563W}. These observation have been carried out with the Very Large Array (VLA) and are characterized by a high angular resolution (6 arcsec and 10 arcsec in the robust and natural weighting, respectively). To compute the HI surface brightness, $\Sigma_{\rm HI}$, we first convolve the 21cm natural-weighted intensity maps, given in Jy beam$^{-1}$ m s$^{-1}$, to the resolution of the worst of the 23 photometric bands used in the SED fitting (the one of SPIRE350, 24 arcsec, see Paper I) using a Gaussian kernel. We used the beam sizes given in Table\,2 of \citet{2008AJ....136.2563W} and Eq.\,1 to obtain the flux in K km s$^{-1}$ and then estimate $\Sigma_{\rm{HI}}$ from Eq.\,5 of \citet{2008AJ....136.2563W} (that does not include a correction for helium). We compute the sensitivity limit from our maps of $\Sigma_{\rm HI}$ at 500 pc resolution and find $\Sigma_{\rm HI,lim}$ $\sim 2$ M$_{\odot}$pc$^{-2}$. The central panel of Fig.\,\ref{fig:0628} shows the distribution of $\Sigma_{\rm HI}$ in NGC0628. As in several other spiral galaxies, the HI is centrally depressed \citep[e.g.][]{2017A&A...605A..18C}, and it extends on radius that are significantly larger that the optical radius \citep{2002A&A...390..829S, 2013MNRAS.433..270W}. The values of M$_{\rm HI}$ within R$_{25}$ are reported in Table \ref{tab:sample}. For the galaxies in our sample, the HI gas-to-stellar mass ratio (M$_{\rm HI}$/M$_{\star}$) within R$_{25}$  varies from 5\% to 60\%.

\subsection{Molecular gas: CO observations}
The molecular gas surface density, $\Sigma_{\rm{H_2}}$, is computed using the $^{12}$CO$(2-1)$ intensity maps from the HERACLES survey \citep[The HERA CO-Line Extragalactic Survey,][]{2009AJ....137.4670L}. These observations were made with the IRAM 30m telescope and have an angular resolution of 11 arcsec. As for $\Sigma_{\rm{HI}}$ we convolve the images using a Gaussian kernel to the resolution of SPIRE350. We estimated $\Sigma_{\rm{H_2}}$ using Eq.\,4 of \citet{2009AJ....137.4670L}, considering a metallicity independent conversion factor X$_{\rm CO}$ (X$_{\rm CO} = {\rm N}({\rm H}_2)/{\rm I}_{\rm CO}$, where N(H$\rm{_2}$) is the H$\rm{_2}$ column density and I$_{\rm CO}$ is the line intensity) equal to $2\cdot 10^{20} \mathrm{\ cm^{-2} (K\ km\ s^{-1})^{-1}}$ \citep[the typical value for disc galaxies, see e.g.][]{2013ARA&A..51..207B}, and a CO line ratio I$_{{\rm CO}(2-1)}/{\rm I}_{{\rm CO}(1-0)}=0.8$ \citep[e.g.][]{2009AJ....137.4670L,2011AJ....142...37S,2015A&A...577A.135C}. We divide by a factor 1.36 that is included in Eq.\,4 of \citet{2009AJ....137.4670L} to remove the helium contribution. In Sec.\,\ref{sec:Results} we show that the results presented here remain true when considering a metallicity-dependent X$_{\rm CO}$ factor, using the X$_{\rm CO} - (12+\log {\rm O/H})$ relation of \citet{2011ApJ...733..101G} and the spatially resolved metallicity measurements collected in DustPedia. The sensitivity limit, computed as the rms of our log$\Sigma_{\rm H_2}$ maps at 500pc resolution, is log$\Sigma_{\rm H_2,\,lim}$ = 0.4 M$_{\odot}$ pc$^{-2}$. For the regions corresponding to a negative flux of $^{12}$CO$(2-1)$, in the $\Sigma_{\rm H_2,\,lim}$ map we replace the value with a randomly generated number between 0 and the sensitivity limit, so that $\Sigma_{\rm H_2}$ can be computed as an upper limit. While this step does not influence our results concerning H$_2$, it allows us to extend the analysis also to the regions where H$_2$ is not detected. The right panel of Fig.\,\ref{fig:0628} shows the distribution of $\Sigma_{\rm H_2}$ in NGC0628. The H$_2$ is centrally concentrated, and mostly below the sensitivity limit for $r>0.5$ R$_{25}$. For the galaxies in our sample, the H$_2$ gas-to-stellar mass ratio (M$_{\rm H_2}$/M$_{\star}$) within R$_{25}$ is almost constant around 11-14\%.

\section{Results}\label{sec:Results}
Before analysing the spatially resolved connection between star formation and gas components, we briefly comment on the integrated properties of the galaxies in our sample. It is worth underlying that, within R$_{25}$, three out of five galaxies have similar amount of neutral and molecular gas (M$_{\rm H_2}$ and M$_{\rm HI}$), within the uncertainties. This is consistent with the results of \citet{2020A&A...633A.100C}, as they find that, within R$_{25}$, galaxies with morphological type T = 4, 5, 6 have M$_{\rm H_2}$/M$_{\rm HI}$ = 0.91, 1 and 1.05 respectively. For NGC5457 and NGC5194 the average value associated to their morphological type does not describe well their gas properties. It is well know that NGC5457 is likely to have experienced a recent event of gas accretion \citep{2013ApJ...762...82M, 2019MNRAS.483.4968V} which can explain the HI rich outer disc and its high total HI mass. A high molecular gas mass fraction, as for NGC5194, is likely the result of tidal stirring by a companion.

\begin{figure*}
	\includegraphics[width=0.99\textwidth]{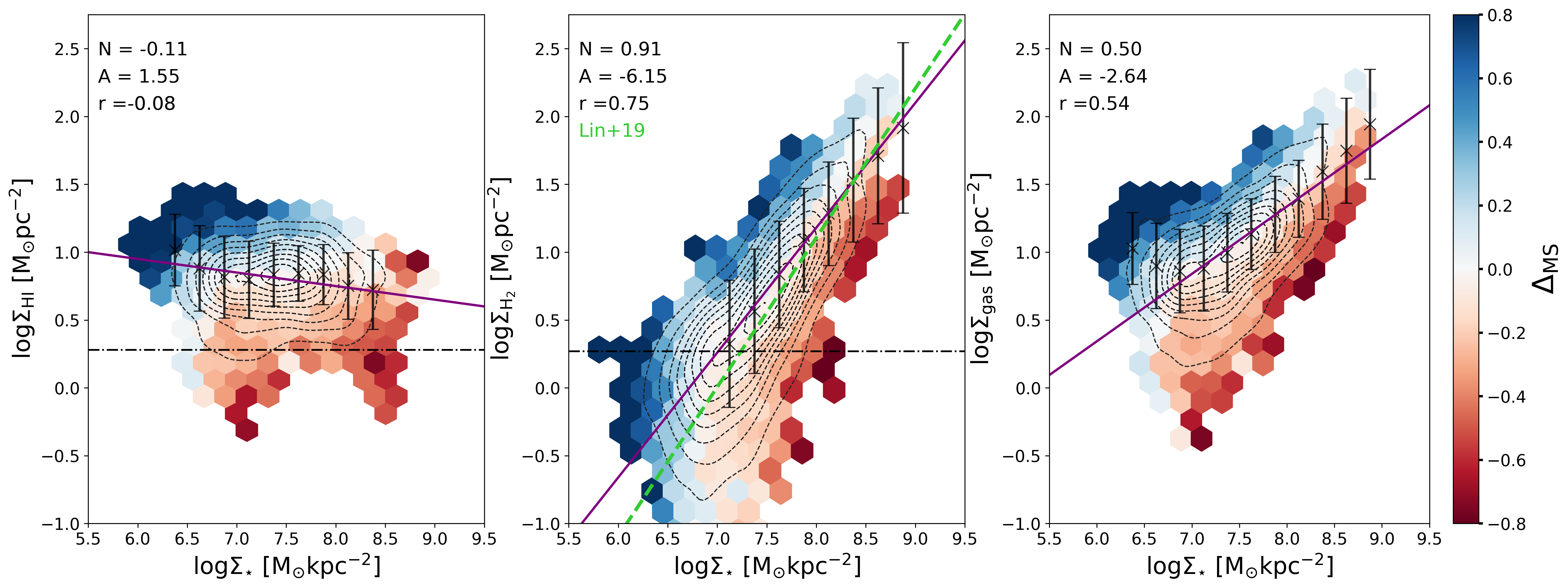}
    \caption{Distributions of the regions in the $\log\Sigma_{\star}$-$\log\Sigma_{\rm HI}$ plane (left panel), $\log\Sigma_{\star}$-$\log\Sigma_{\rm H_2}$ plane (central panel) and $\log\Sigma_{\star}$-$\log\Sigma_{\rm gas}$ plane (right panel). Each hexagonal bin the the plane has been colour coded according to the average value of $\Delta_{\rm MS}$, as in Fig. \ref{fig:bigiel}. The purple solid lines are the best fit relations obtained by fitting the average values of log$\Sigma_{\rm HI,H_2,gas}$ in bins of 
    $\log\Sigma_{\star}$, as marked with black crosses; the slope (N) and intercept (A) of the best fit are written in the panels, together with the Spearman correlation coefficient \rm{r}. The green dashed line in the central panel is the MGMS of \citet{2019ApJ...884L..33L}, re-scaled to a Chabrier IMF.}
    \label{fig:mgms}
\end{figure*}

\subsection{Dependency of the SFR on gas}
\label{sec:ks}
With the data set in our hands, we first investigate the spatially resolved relations between the SFR and the different gas phases, by analysing how $\Sigma_{\rm SFR}$ relates to $\Sigma_{\rm HI}$, $\Sigma_{\rm H_2}$, and $\Sigma_{\rm gas}$. This last quantity is computed as the sum of the neutral and molecular component for all the regions where the $\Sigma_{\rm H_2}$ is above the sensitivity limit, while it is equal to $\Sigma_{\rm H_I}$ otherwise, and it is thus a lower limit. The results of this exercise are shown in Fig.\,\ref{fig:bigiel}.

As expected, no significant correlation is found between $\Sigma_{\rm SFR}$ and $\Sigma_{\rm HI}$ (Spearman correlation coefficient r = 0.26), while tight correlations are present between $\Sigma_{\rm SFR}$ and $\Sigma_{\rm H_2}$, and $\Sigma_{\rm SFR}$ and $\Sigma_{\rm gas}$ (r = 0.76), confirming several results in the literature \citep[e.g.][]{2008AJ....136.2846B,2020A&A...634A..24K}. In Fig.\,\ref{fig:bigiel} we add the information on how regions located at different distance from the spatially resolved MS populate the log$\Sigma_{\rm SFR}$ - log$\Sigma_{\rm HI/H_2/gas}$ plane. To do so, we divide the log$\Sigma_{\rm SFR}$ - log$\Sigma_{\rm HI/H_2/gas}$ planes in bins colour coded according to the median value of $\Delta_{\rm MS}$ in each bin. We observe that regions above the MS are found in correspondence to the highest $\Sigma_{\rm HI}$, but span a wide range of $\Sigma_{\rm H_2}$ values. Analogously, regions located below the MS are preferentially found at lower $\Sigma_{\rm HI}$, while are characterized by $\Sigma_{\rm H_2}$ spanning the whole range of possible values. For log$\Sigma_{\rm SFR} > -2$ the trend between log$\Sigma_{\rm HI}$ and $\Delta_{\rm MS}$ is less evident; this is due to the fact that we are not well sampling the region below the MS, as shown in Fig.\,5 of Paper I, and confirmed by the decrease of the scatter of the spatially resolved MS at the higher stellar surface densities. Analogously, the points with log$\Sigma_{\rm SFR} < -3$ are mostly found on the MS or below it, thus hiding a possible trend at the lowest SFRs. When considering the total gas, we see that regions closer to the relation that describes local Ultra-Luminous IR Galaxies (ULIRGs) and sub-mm galaxies \citep[dashed black lines, taken from][]{2010ApJ...714L.118D} are the ones located above the relation. On the other hand, the general behaviour between the distance from the best fit relation and the distance from the MS is less regular than in the case of molecular gas. Indeed, the central panel of Fig.\,\ref{fig:bigiel} suggests that the spatially resolved MS is intrinsically linked to the molecular gas - SFR relation, as MS regions (in white) fall very consistently along the log$\Sigma_{\rm SFR}$ - log$\Sigma_{\rm H_2}$ relation. Regions that populate the upper (lower) envelope of the molecular KS law are also found in the upper (lower) envelope of the spatially resolved MS relation. This is qualitatively consistent with what found by \cite{2020MNRAS.493L..39E} when analysing the molecular gas - SFR relation exploiting ALMA and MaNGA data, on kpc scales. 

By fitting the average values of log$\Sigma_{\rm SFR}$ in bins of log$\Sigma_{\rm H_2}$ and log$\Sigma_{\rm gas}$ (both above the sensitivity limit of log$\Sigma_{\rm H_2}$) we find the following scaling relations (slopes N and intercepts A are also written in the corresponding panels):
\begin{equation}\label{ks}
    \log \Sigma_{\rm SFR} = 0.80(\pm0.12)\cdot \log \Sigma_{\rm H_2}-2.97(\pm0.85)
\end{equation}
and
\begin{equation}\label{eq:SFR-gas_ODR}
    \log \Sigma_{\rm SFR} = 1.21(\pm0.18)\cdot \log \Sigma_{\rm gas}-3.74(\pm1.15)
\end{equation}

 The two correlations have equal strength according to the Spearman coefficient, and their scatter is in both cases smaller than the one of the spatially resolved MS: 0.19 for the log$\Sigma_{\rm H_2}$ - log$\Sigma_{\rm SFR}$ relation and 0.17 for the log$\Sigma_{\rm gas}$ - log$\Sigma_{\rm SFR}$. The relation between log$\Sigma_{\rm SFR}$ and log$\Sigma_{\rm H_2}$ is sub-linear, but becomes linear when the ODR fitting is applied. We retrieve a molecular t$_{\rm depl}$ that varies between 1.6 and 3 Gyr. These results are in agreement with the typical t$_{\rm depl}$ in normal spiral galaxies \citep[see, e.g.][]{2008AJ....136.2846B,2011MNRAS.415...61S,2013AJ....146...19L,2015A&A...577A.135C}. It is worth noticing that the slope of the spatially resolved KS relation obtained from the molecular and total gas are consistent, within errors, with the integrated relations \citep[e.g.][]{2010MNRAS.407.2091G,2012ARA&A..50..531K,2013ApJ...768...74T,2019ApJ...872...16D,2019A&A...622A.105F}, similar to what is found when comparing the spatially resolved and integrated MS relation. The scatter of the integrated relation is instead significantly higher \citep[e.g. 0.28 in ][]{2019ApJ...872...16D}.

Finally, we underline here that the trends observed in Fig.\,\ref{fig:bigiel} are not driven by one or few of the galaxies in our sample, but by and large are common to all five galaxies in our sample. Slope and intercept of the $\log\Sigma_{\rm{H_2}}$-$\log\Sigma_{\rm {SFR}}$ and $\log\Sigma_{\rm{gas}}$-$\log\Sigma_{\rm {SFR}}$ relations are summarised in Table \ref{tab:metallicity}.

\begin{figure*}
	\includegraphics[width=0.99\textwidth]{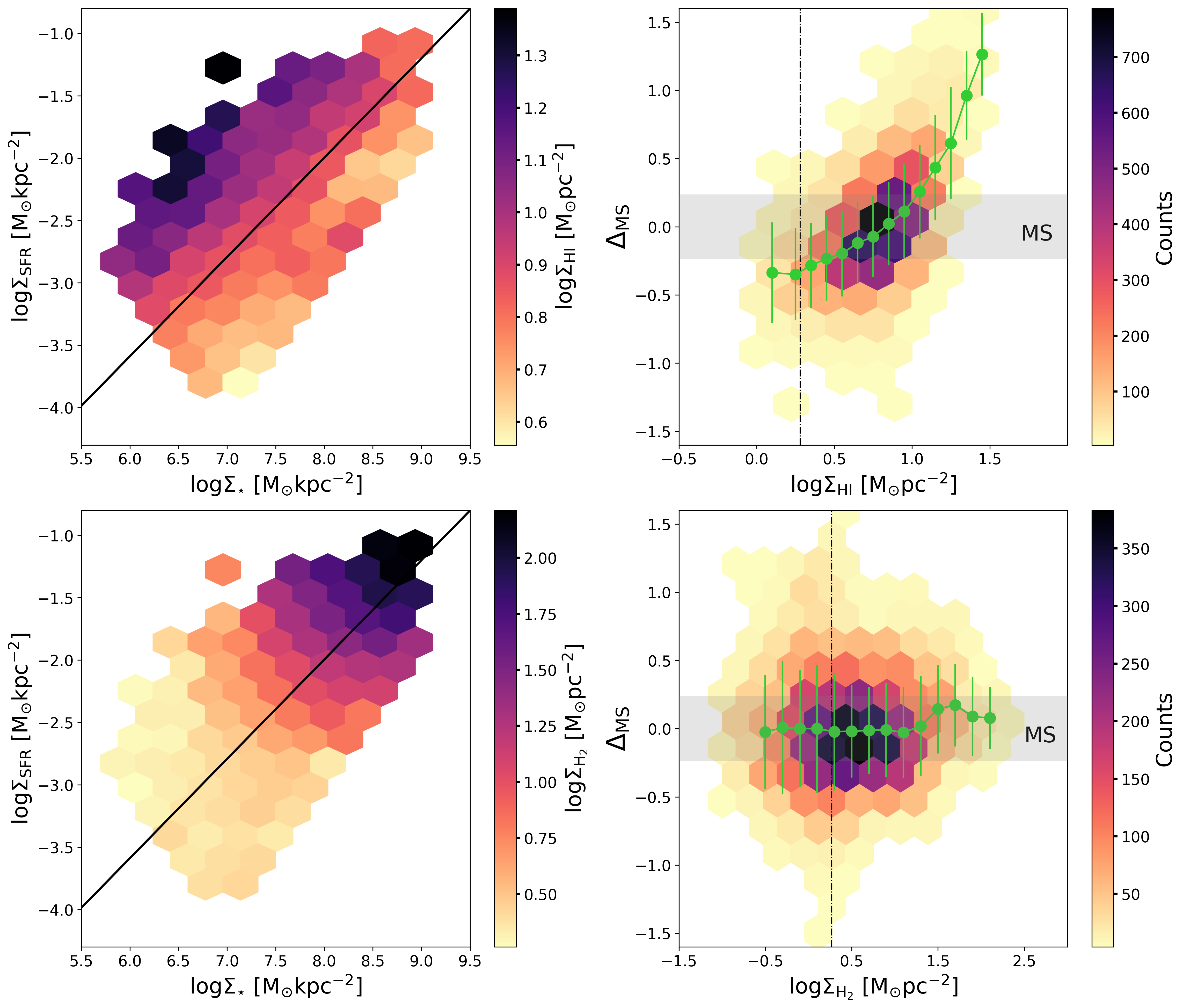}
    \caption{HI and H$_2$ in the log$\Sigma_{\star}$ - log$\Sigma_{\rm SFR}$ plane. \textit{Top-left panel:}  log$\Sigma_{\star}$ - log$\Sigma_{\rm SFR}$ plane color coded as a function of the average value of log$\Sigma_{\rm HI}$ in each bin. The black sold line marks the location of the spatially resolved MS relation. \textit{Top-right panel:} $\Delta_{\rm MS}$,  as a function of log$\Sigma_{\rm HI}$. The average values of  $\Delta_{\rm MS}$ computed in bins of log$\Sigma_{\rm HI}$ are shown in green. Each bin is colour coded as a function of the number of cells that it contains. \textit{Bottom-left panel:}  log$\Sigma_{\star}$ - log$\Sigma_{\rm SFR}$ plane color coded as a function of the average value of log$\Sigma_{\rm H_2}$ in each bin. The black sold line marks the location of the spatially resolved MS relation. \textit{Bottom-right panel:} $\Delta_{\rm MS}$,  as a function of log$\Sigma_{\rm H_2}$. The average values of  $\Delta_{\rm MS}$ computed in bins of log$\Sigma_{\rm H_2}$ are shown in green. Each bin is colour coded as a function of the number of cells that it contains. In the right panels, the grey shaded area marks the MS region.}
    \label{fig:fig4}
\end{figure*}

\subsection{Dependency of gas distribution on stellar mass}
\label{sec:mass_gas}

As the trends shown in Fig.\,\ref{fig:bigiel} with $\Delta_{\rm MS}$ are related to variations of the gas content with M$_{\star}$, we show in Fig.\,\ref{fig:mgms} how the surface densities of neutral,molecular and total gas vary as a function of $\Sigma_{\star}$. The left panel of Fig.\,\ref{fig:mgms} shows that a very weak anti-correlation is found between log$\Sigma_{\star}$ and log$\Sigma_{\rm HI}$. Indeed, when fitting the average values of $\log\Sigma_{\rm HI}$ in bins of $\log\Sigma_{\star}$, the slope of the correlation is $-0.11\pm0.07$. We stress that every galaxy shows a trend of decreasing $\Sigma_{\rm HI}$ towards the central regions, as in the central region the high pressure favours the HI to H$_2$ transition and most of the gas is in molecular form, but such a decrease can be more or less pronounced from galaxy to galaxy, and does not follow a universal behaviour. Starbursting regions are preferentially located at $r>0.5$ R$_{25}$, where the surface density of stars falls below $10^7$ M$\odot$ pc$^{-2}$, and are generally found along the spiral arms. 

The relation between $\log\Sigma_{\star}$ and $\log\Sigma_{\rm H_2}$, shown in the central panel of Fig.\,\ref{fig:mgms} is consistently common to all five galaxies and gives birth to a very tight correlation, the MGMS \citep{2019ApJ...884L..33L}. The MGMS (re-scaled to a Chabrier IMF) of \citet{2019ApJ...884L..33L} is indicated with a green dashed line in the central panel of Fig.\,\ref{fig:mgms} and has a slope of 1.1. To obtain the slope of our MGMS relation, we fit the average values of $\log\Sigma_{\rm H_2}$ in bins of $\log\Sigma_{\star}$, restricting the analysis to stellar surface densities where the average value of $\log\Sigma_{\rm H_2}$ is above the sensitivity limit. We find:
\begin{equation}
      \log \Sigma_{\rm H_2} = 0.91(\pm0.29)\cdot \log \Sigma_{\star}-6.15(\pm2.11)
\label{mgms}
\end{equation}

This relation has a scatter of 0.22 dex, similar to that obtained for the MS (0.23dex) and the Spearman coefficient is the same as for the $\log\Sigma_{\rm{H_2/gas}}$-$\log\Sigma_{\rm {SFR}}$ relations. We find a slope that is consistent within the error with the one of \citet{2019ApJ...884L..33L}, that is 1.1. The different slopes can be ascribed to different fitting procedures; indeed, if we follow the ODR method, we also retrieve a super-linear slope. Regions with the largest sSFR are located in the upper envelope of this relation; in other words, cells located above the spatially resolved MS are also located above the $\log\Sigma_{\star}$ - $\log\Sigma_{\rm H_2}$ relation. This trend, outlined also in \cite{2020MNRAS.493L..39E}, is here as significant as the one visible in Fig.\,\ref{fig:bigiel} for the molecular KS relation.

A correlation is also apparent between $\log\Sigma_{\star}$ and $\log\Sigma_{\rm gas}$, that is the combination of the two behaviours seen in the left and central panel of Fig.\,\ref{fig:mgms}. At low $\Sigma_{\star}$ ($\log\Sigma_{\star}\lesssim$ 7), the HI tends to dominate over H$_2$, and the scatter of the relation is larger, while it is slightly narrower at large $\Sigma_{\star}$. This combined behaviour results in a slope of $0.50\pm0.14$ and an intercept of $-2.64\pm0.97$, and a Spearman coefficient of 0.54. Slope and intercept of the $\log\Sigma_{\star}$-$\log\Sigma_{\rm gas}$ and  $\log\Sigma_{\star}$-$\log\Sigma_{\rm H_2}$ relations are summarised in Table \ref{tab:metallicity}.

\subsection{HI and H$_2$ in the $\Sigma_{\star}$-$\Sigma_{\rm SFR}$ plane} 
To further analyse the link between the atomic, molecular and total gas and the star formation properties of a region, we show in Fig.\,\ref{fig:fig4} how log$\Sigma_{\rm HI}$ and log$\Sigma_{\rm H_2}$ vary across the log$\Sigma_{\rm \star}$ - log$\Sigma_{\rm SFR}$ plane. In the two left panels of Fig.\,\ref{fig:fig4} we show the log$\Sigma_{\rm \star}$ - log$\Sigma_{\rm SFR}$ plane color coded as a function of the average log$\Sigma_{\rm HI}$ (top) and log$\Sigma_{\rm H_2}$ (bottom) values in each bin. The spatially resolved MS is indicated with the black solid line. To compute the average value of log$\Sigma_{\rm H_2}$ we also used values below the sensitivity limit, therefore it is important to emphasise that H$_2$ is detected only for $\Sigma_{\rm H_2}$ above $\sim$ 3 M$_{\odot}$pc$^{-2}$  which requires a $\Sigma_{\star}$ higher than 10$^7$M$_{\odot}$kpc$^{-2}$. On the other hand H$_2$ column densities below this threshold value are hardly self-shielded and quite rare \citep[e.g.][]{2014ApJ...790...10S}.

Interestingly, along the MS relation the average value of log$\Sigma_{\rm HI}$ is fairly constant and equal to about 7 M$_{\odot}$pc$^{-2}$  which corresponds to an HI column density of about 9$\cdot$10$^{20}$ cm$^{-2}$. This indicates that for cells on the MS the radiation field is strong enough to partially dissociate H$_2$, and large amounts of dust rich HI gas prevent further H$_2$  dissociation \citep{2014ApJ...790...10S}. As expected, $\Sigma_{\rm H_2}$ increases with increasing M$_{\star}$, following the MGMS relation, suggesting that the gravity dominated by stars compresses and enhances the ISM volume density, thus favoring the formation of molecules. The upper envelope of the MS is populated by cells that, on average, have larger HI surface densities than counterparts located on the relation and below it. Focusing on H$_2$, we can see that no variations in log$\Sigma_{\rm H_2}$ are visible perpendicular to the MS relation, while a weak trend on increasing log$\Sigma_{\rm H_2}$ towards higher log$\Sigma_{\rm SFR}$ can be seen at fixed stellar surface density. For example, at log$\Sigma_{\star}$ = 7.5 the average value of log$\Sigma_{\rm H_2}$ on the MS is 0.8 M$\odot$ pc$^{-2}$, and increases to $\sim$1.35 M$\odot$ pc$^{-2}$ 0.8 dex above the relation. Following Eq.\,\ref{ks}, this variation in log$\Sigma_{\rm H_2}$ would correspond to a difference in log$\Sigma_{\rm SFR}$ of $\sim$0.5 dex, implying that the increase in H$_2$ seen at fixed stellar surface densities is not sufficient alone in setting large SFR. In the right panels of Fig.\,\ref{fig:fig4} we show the distance from the spatially resolved MS plotted as a function of log$\Sigma_{\rm HI}$ (top) and log$\Sigma_{\rm H_2}$ (bottom). As indicated qualitatively from Fig.\,\ref{fig:bigiel}, we observe a correlation between log$\Sigma_{\rm HI}$ and $\Delta_{\rm MS}$, for which regions located above the MS are characterized by the largest HI surface brightness, while regions below the MS correspond to cells with low log$\Sigma_{\rm HI}$. We do not observe any correlation between log$\Sigma_{\rm H_2}$ and $\Delta_{\rm MS}$. This is the combination of the two previous results: 1) the molecular KS relation is populated by regions on the MS, and 2) the existence of the MGMS. This suggests that the absolute quantity of molecular gas in a region (or, equivalently its surface density) is not related to the sSFR of the region itself.  

\subsection{Dependency of the results on metallicity}

\begin{figure}
	\includegraphics[width=0.49\textwidth]{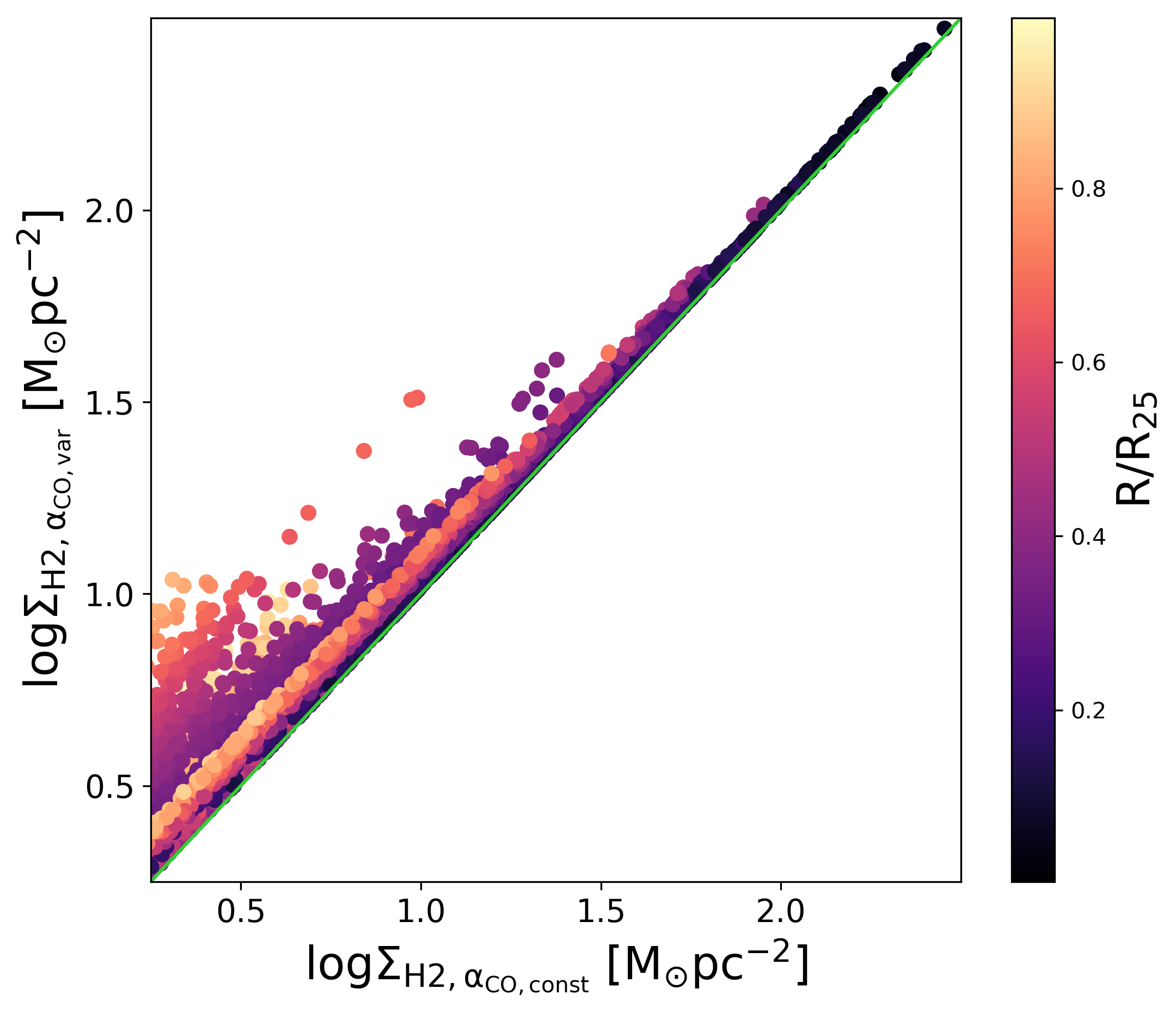}
	\caption{Comparison between the spatially resolved molecular gas mass computed with a constant $\alpha_{\rm CO}$ factor and a metallicity dependent one (here for clarity we use the metallicity based on the O3N2 measurement; no significant differences are found when using the metallicity based on the N2 data). The points are colour coded as a function of their R/R$_{25}$ value. The green solid line marks the 1 to 1 relation.}
    \label{fig:met}
\end{figure}

A possible source of uncertainty in this work is the dependency of the conversion factor between CO and H$_2$ on gas-phase metallicity ($\alpha_{\rm CO}$). Indeed, several works have shown that $\alpha_{\rm CO}$ varies strongly as a function of the metallicity \citep[e.g.][]{2013ARA&A..51..207B}, and strong metallicity gradients have been found in some of the galaxies in this sample, as well as in larger samples of local star forming galaxies \citep[][]{2014MNRAS.444.3894H, 2018ApJ...865..117C, 2019MNRAS.483.4968V}, reaching a factor of 10 within the optical radius. From an integrated perspective, instead, \citet{2015ApJ...800...20G} find that $\alpha_{\rm CO}$ varies little within $\pm$0.6 dex of the MS (thus for the large majority of the cells in this work). Nevertheless, such variations in metallicity need to be addressed properly in order to avoid biased interpretations of spatially resolved results. From the DustPedia archive, we download the table containing all the metallicity measurements available in literature \citep{2019A&A...623A...5D} and obtain in regions within R$_{25}$ of the five galaxies in our sample. For NGC5457 there are 280 estimates of metallicity within the optical radius, while for NGC6946 only 14 are available. In particular, we make use of the metallicities computed exploiting the N2 and O3N2 calibrations of \citet{2004MNRAS.348L..59P}\footnote{We refer the reader to \citet{2020A&A...633A.100C} and \citet{2019A&A...623A...5D} for details of the metallicity calibration.}. In particular, N2 is defined as $\log([\ion{N}{ii}_{6583A} / H_{\alpha})$ and O3N2 as $\log[([\ion{O}{iii}]_{5007A} / H_{\beta}) / ([\ion{N}{ii}]_{6583A} / H_{\alpha})]$. With the conversion relations of \citet{2004AJ....127.2002K} we obtain the metallicities in the \citet{2002MNRAS.330...69D} calibration. For each galaxy we then build a 1D metallicity profile by fitting the different measurements. We use the 1D metallicity profile to obtain a 2D map of the metallicity dependent $\alpha_{\rm CO}$ factor exploiting the relation of \citet{2012ApJ...746...69G}, that is obtained by fitting the z$\sim$0 points of \citet{2011ApJ...737...12L} with z$>$1 ones collected in \citet{2012ApJ...746...69G}:
\begin{equation}
    \log \alpha_{\rm CO} = -1.3\cdot (12+\log(O/H))_{\rm Denicolo02} + 12
\end{equation}
With the 2D map of $\log \alpha_{\rm CO}$ we then estimate $\Sigma_{\rm H_2}$. Fig.\,\ref{fig:met} shows a comparison of the molecular gas surface density computed considering a constant $\alpha_{\rm CO}$ and the metallicity dependent one (in particular, the one obtained with the O3N2 calibration, but no significant differences are found when considering the N2 one). The highest scatter corresponds to small $\Sigma_{\rm H_2}$, located in the outskirts of the optical disc, where the metallicity is, on average, smaller than in the centre. The source characterized by the largest scatter is NGC5457, that is also the one with the strongest metallicity gradient \citep{2019MNRAS.483.4968V}. We note, nevertheless, that within 0.5 R$_{25}$ (that is, in first approximation, the distance within which the estimate of $\Sigma_{\rm H_2}$ is above the sensitivity limit), the maximum difference between the two estimates is around 0.2 dex. 

\begin{figure*}
	\includegraphics[width=0.98\textwidth]{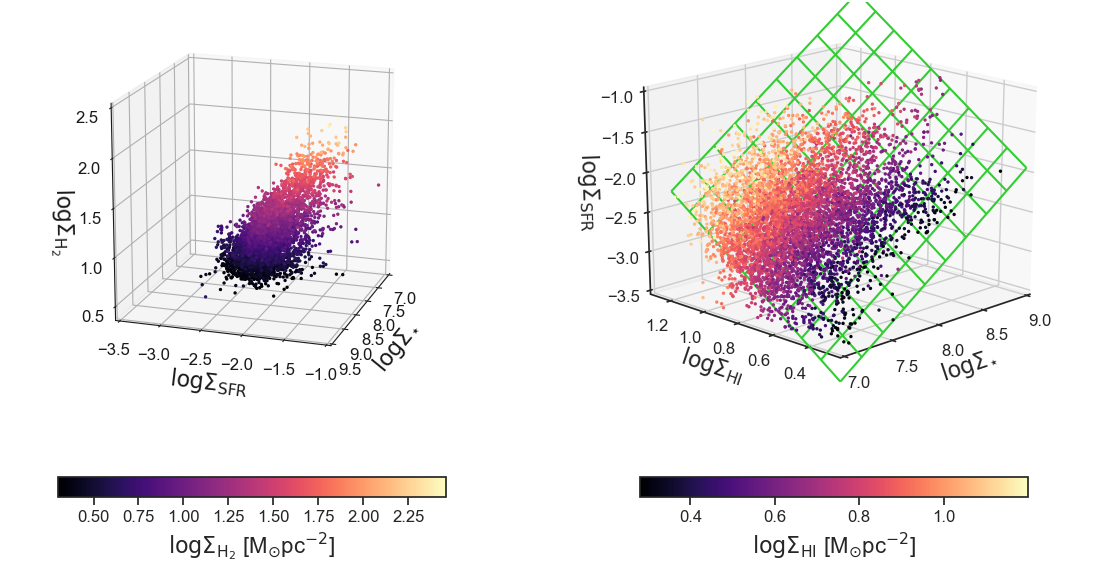}
    \caption{ \textit{Left panel}: distribution of the regions having an estimate of $\Sigma_{\rm H_2}$ above the sensitivity limit in the $\log\Sigma_{\star}$ - $\log\Sigma_{\rm SFR}$ - $\log\Sigma_{\rm H_2}$ plane. Each point is colour coded as a function of $\log\Sigma_{\rm H_2}$. Four projections of this space at different azimuthal angles are shown in Fig.\,\ref{fig:sausage}. \textit{Right panel}: distribution of the regions having an estimate of $\Sigma_{\rm HI}$ above the sensitivity limit in the $\log\Sigma_{\star}$ - $\log\Sigma_{\rm SFR}$ - $\log\Sigma_{\rm HI}$ plane. Each point is colour coded as a function of $\log\Sigma_{\rm HI}$. The best fit plane is indicated by the green grid. Four projections of this plane, described by Eq.\,\ref{eq:plane}, at different azimuthal angles are shown in Fig.\,\ref{fig:plane}.}
    \label{fig:3d}
\end{figure*}

\begin{table*}
	\centering
	\caption{Slope, intercept and scatter of the following relations: molecular KS law, $\log\Sigma_{\rm gas}$ - $\log\Sigma_{\rm SFR}$, MGMS and  $\log\Sigma_{\star}$ - $\log\Sigma_{\rm gas}$. We list the best fit parameters for $\log\Sigma_{\rm H_2}$ computed using a constant XCO factor and a metallicity dependent XCO obtained from estimates of the metallicity that use: 1) the N2 index, and 2) the O3N2 index.}
	\label{tab:metallicity}
	\begin{tabular}{lccccccccccc}
		\hline
		Correlation     &  \multicolumn{3}{c}{Slope} &  \multicolumn{3}{c}{Intercept} & \multicolumn{3}{c}{Scatter} \\
		                & const $\alpha_{CO}$ & O3N2 & N2 & const $\alpha_{CO}$ & O3N2 & N2 & const $\alpha_{CO}$ & O3N2 & N2 \\
		\hline
        $\log\Sigma_{\rm H_2}$ - $\log\Sigma_{\rm SFR}$  & 0.80$\pm$0.12 & 0.83$\pm$0.12 &  0.83$\pm$0.11 & -2.97$\pm$0.87 & -2.97$\pm$0.93 & -2.94$\pm$0.94 & 0.19 & 0.20 & 0.19\\
        $\log\Sigma_{\rm gas}$ - $\log\Sigma_{\rm SFR}$  & 1.21$\pm$0.18 &  1.24$\pm$0.16 & 1.25$\pm$0.18  & -3.74$\pm$1.15 &  -3.74$\pm$1.15 &  -3.73$\pm$1.14& 0.17 & 0.18 & 0.18\\
        $\log\Sigma_{\star}$ - $\log\Sigma_{\rm H_2}$  & 0.91$\pm$0.29 & 0.84$\pm$0.29 &  0.87$\pm$0.27 &-6.15$\pm$2.11 & -5.66$\pm$2.12 & -5.93$\pm$2.07 & 0.22 & 0.24 & 0.23\\
        $\log\Sigma_{\star}$ - $\log\Sigma_{\rm gas}$  & 0.50$\pm$0.14 & 0.49$\pm$0.14 &  0.49$\pm$0.13 & -2.64$\pm$0.97& -2.62$\pm$1.08 & -2.72$\pm$1.05 & 0.24 & 0.25 & 0.25\\
		\hline
	\end{tabular}
\end{table*}

We repeat the previously shown analysis considering $\Sigma_{\rm H_2}$ estimated with the metallicity dependent $\alpha_{\rm CO}$. In Tab.\,\ref{tab:metallicity} we report the slopes, intercepts and scatter of the various relations discussed above:  $\log\Sigma_{\rm H_2}$-$\log\Sigma_{\rm {SFR}}$, $\log\Sigma_{\rm{gas}}$-$\log\Sigma_{\rm {SFR}}$, $\log\Sigma_{\star}$-$\log\Sigma_{\rm {H_2}}$, and $\log\Sigma_{\star}$-$\log\Sigma_{\rm {gas}}$. With respect to the case of constant $\alpha_{\rm CO}$, the slope of molecular KS relation slightly increases but not significantly given the errors on the estimates. The slope of the total gas KS law decreases to reach values closer to 1, but again this decrease is not significant when taking the errors into account. Similarly, the variations in slope and intercept of the  $\log\Sigma_{\star}$ - $\log\Sigma_{\rm{H_2, gas}}$ relations do not vary significantly with respect to the values found in Sec.\,\ref{sec:mass_gas}. The scatter of the four relations slightly increases. In particular, the scatter of the $\log\Sigma_{\star}$ - $\log\Sigma_{\rm{H_2, gas}}$ relations is comparable or larger than the one of the spatially resolved MS. This is expected, as we are adding several sources of uncertainty: the conversion between different metallicity calibrations, the correlation between $\alpha_{\rm CO}$ and metallicity, as well as the fact that we are averaging the metallicity to obtain a 1D profile. In first approximation, this exercise reveals that the result in this paper are robust against variations of the $\alpha_{\rm CO}$ factor as a function of metallicity, that is the main and most studied dependence of $\alpha_{\rm CO}$ on physical/galaxy properties (e.g., gas temperature and abundance, optical depth, cloud structure, cosmic ray density, and UV radiation field, in addition to the metallicity). This happens because the central metallicities for the galaxies in our sample are similar, and the molecular gas maps are not deep enough to reach the outermost regions where the metallicity decrease with respect to the central value.

\begin{figure*}
	\includegraphics[width=0.99\textwidth]{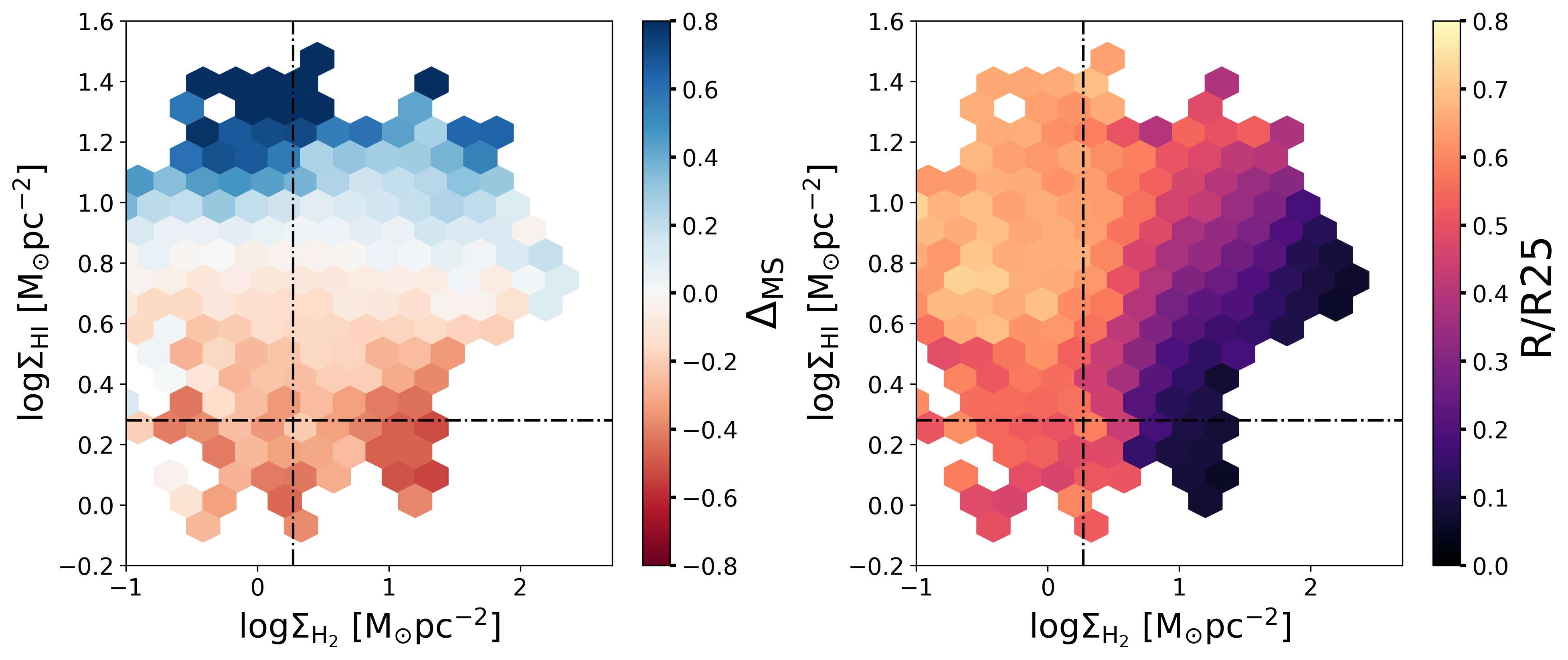}
	 \caption{The log$\Sigma_{\rm H_2}$ - log$\Sigma_{\rm HI}$ plane, colour coded as a function of $\Delta_{\rm MS}$ (left panel) and R/R$_{25}$ (right panel). Sensitivity limits are marked with dotted-dashed lines.}
    \label{fig:hih2}
\end{figure*}

\section{Discussion}\label{sec:Discussion}

\subsection{The origin of the main sequence and its scatter}
The spatially resolved MS of galaxies constitutes the building block of the integrated MS relation of star forming galaxies so deeply analysed in literature to understand the star formation processes and the quenching mechanisms. When analysing the molecular gas component of nearby galaxies we find tighter relations than the MS itself, and that may be at the physical origin of it: the KS law (\ref{ks}) and the MGMS (\ref{mgms}). By combining Equations \ref{ks} and \ref{mgms}, we obtain a spatially resolved MS in the form:  
\begin{equation}\label{eq:spres_MS}
    \log\Sigma_{\rm SFR}  = 0.73\cdot \log\Sigma_{\star}-7.89
\end{equation}

that is, as expected, consistent with the spatially resolved MS relation characterising our sample ($\log\Sigma_{\rm SFR}$ = 0.76$\log\Sigma_{\star}$ - 8.15, see Sec. \ref{2e1}). We find that the $\Sigma_{\rm H_2}$ - $\Sigma_{\rm SFR}$ relation is the tightest one of the three, with a scatter of 0.19 dex, followed by the $\Sigma_{\star}$ - $\Sigma_{\rm H_2}$ relation, 0.22 dex, similar to the scatter of the spatially resolved MS (0.23 dex). Contrary to \citet{2019ApJ...884L..33L} we find that the scatter of the spatially resolved MS is significantly smaller than the quadratic sum of the scatters of the $\Sigma_{\star}$ - $\Sigma_{\rm H_2}$ and $\Sigma_{\rm H_2}$ - $\Sigma_{\rm SFR}$ relations, indicating that the scatters of these relations are not independent. Indeed, regions located in the upper (lower) envelope of the molecular KS relation also populate the upper (lower) envelope of the MGMS. To further investigate the connection between the three relations (KS, MS, MGMS), we plot in the left panel of Fig.\,\ref{fig:3d} how regions with an estimate of $\Sigma_{\rm H_2}$ above the sensitivity limit populate the 3D space made by $\log\Sigma_{\star}$, $\log\Sigma_{\rm SFR}$ and $\log\Sigma_{\rm H_2}$ (four different projections of this space are shown in Fig.\,\ref{fig:sausage} in Appendix \ref{sec:a1}). The variables define a 3D relation, as found by \citet{2019ApJ...884L..33L}. This is expected, as we find no dependency of $\Delta_{\rm MS}$ on $\Sigma_{\rm H_2}$ (see Fig.\,\ref{fig:fig4}). On the other hand, the analysis of $\Sigma_{\rm HI}$ in the $\log\Sigma_{\star}$-$\log\Sigma_{\rm SFR}$ plane revealed that the spatially resolved MS scatter seems to be connected to the presence of neutral gas. Indeed, when analysing how regions populate the 3D space formed by $\log\Sigma_{\star}$, $\log\Sigma_{\rm SFR}$ and $\log\Sigma_{\rm HI}$, we find that they identify a plane, as shown in the right panel of Fig.\,\ref{fig:3d} (four different projections of this plane are shown in Fig.\,\ref{fig:plane} in Appendix \ref{sec:a1}). The equation of the plane that minimises the perpendicular distance of the points can be written as: 
\begin{equation}
    \log\Sigma_{\rm SFR} = 0.97\log\Sigma_{\star} + 1.99\log\Sigma_{\rm{HI}} - 11.11
    \label{eq:plane}
\end{equation}

The relation expressed by Eq.\,\ref{eq:plane} has a scatter of 0.14 dex, significantly smaller than the one of the spatially resolved MS relation. Interestingly, when the dependency of the SFR of HI surface densities is taken into account, the relation between $\log\Sigma_{\rm SFR}$ and $\log\Sigma_{\star}$ becomes closer to linear. To better understand the origin of the relation expressed by Eq.\,\ref{eq:plane}, in Fig.\,\ref{fig:hih2} we show the log$\Sigma_{\rm H_2}$ - log$\Sigma_{\rm HI}$ plane colour coded as a function of $\Delta_{\rm MS}$ (left panel) and $r$/R$_{25}$ (right panel). From Fig.\,\ref{fig:hih2} we can appreciate that regions with $\Sigma_{\rm HI}$ higher than 10 M$\odot$ pc$^{-2}$ \citep[that is the typical value for the the HI to H2 transition,][]{2008AJ....136.2846B,2008AJ....136.2782L,2012ApJ...748...75L,2014ApJ...784...80L} correspond to the upper envelope of the spatially resolved MS and, on average, are preferentially located between 0.3 and 0.8 R$_{25}$. These regions span a wide range of log$\Sigma_{\rm H_2}$ values: 1) up to log$\Sigma_{\rm H_2}$ = 2M$_\odot$pc$^{-2}$ for 0.3R$_{25}<r<$0.6 R$_{25}$, and 2) below the sensitivity limit for $r >$0.6 R$_{25}$. In the first case, the stellar surface density and average SFR are moderately high: high HI surface densities could be partially due to H$_2$ dissociation and are needed (together with dust) to prevent the further dissociation of the molecular gas by the intense radiation field. In the outer part of the optical disc, the SFRs are on average lower than in the inner disc, and so is the stellar surface density; nevertheless, the SFRs above the MS reach values that are comparable to SFRs on the MS in the inner part of the disc.

\begin{figure}
	\includegraphics[width=0.47\textwidth]{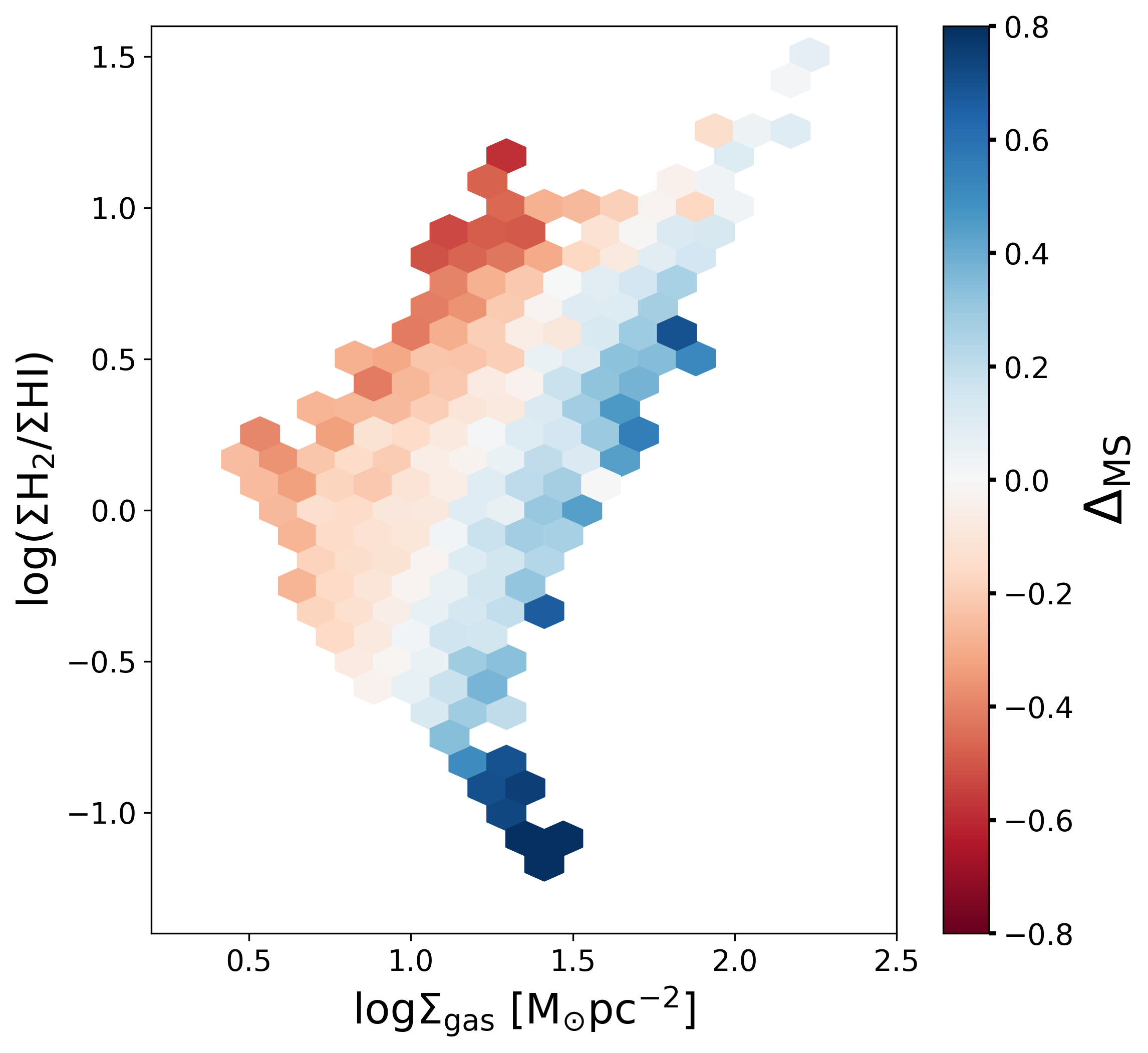}
	 \caption{The $\Sigma_{\rm H_2}$/$\Sigma_{\rm HI}$ ratio as a function of $\Sigma_{\rm gas}$. The cells have been colour coded as a function of the average $\Delta_{\rm MS}$.}
    \label{fig:alvio}
\end{figure}

Most likely, once the molecular gas gives birth to new stars, it
is partly disrupted by the resulting intense UV radiation from hot massive stars. To illustrate this effect, in Fig.\,\ref{fig:alvio} we
plot the $\Sigma_{\rm H_2}$/$\Sigma_{\rm HI}$ ratio as a function of $\Sigma_{\rm gas}$. As expected, for higher $\Sigma_{\rm gas}$ this ratio increases: H$_2$ becomes progressively more dominant over HI as the chemical equilibrium shifts in favor of the molecular phase with increasing gas pressure. However, the correlation is quite broad, and the origin of the spread at fixed $\Sigma_{\rm gas}$ becomes evident once the cells are color coded as a function of $\Delta_{\rm MS}$. Going from below to above the MS relation, $\Sigma_{\rm H_2}$/$\Sigma_{\rm HI}$ steadily decreases. We interpret this trend as evidence that stellar feedback (radiative and from supernova shocks) has the effect of partially dissociating the H$_2$ molecules in regions of intense star formation. The sheer size of this trend is worth emphasizing, as the H$_2$/HI ratio drops by about a factor of 100 when going from extreme sub-MS to super-MS (starbursting) cells. Such a wide range is indeed expected by theory, for a wide variation of the intensity of the UV radiation field \citetext{cf. Figure 7 in \citealp{2014ApJ...790...10S}, see also \citealp{2020arXiv200306245T}}.

We notice that a correlation between the HI abundance and the distance from the MS has been recently reported by \citet{2020arXiv200101970W} when analysing integrated galaxy properties, i.e., galaxies above the MS are more HI-rich than those below. They interpret this trend in terms of HI being an intermediate step (between the ionized and the molecular phase) in fueling star formation in galaxies, but do not consider HI as a product of molecular dissociation. Our results instead indicate that  HI is also a product of star formation and its surface density is extremely sensitive to the local UV radiation field.

\subsection{Star Formation Efficiency vs Gas Fraction}

The data set in our hands gives us the possibility to analyse the role of SFE and $f_{\rm gas}$ in setting the sSFR of a region, thus deciding its location with respect to the spatially resolved MS relation. As in literature the gas fraction and SFE estimates may or may not include the contribution of neutral gas, depending on how the gas mass is measured, we define the molecular, atomic and total SFE: SFE$_{\rm mol}$ = SFR/M$_{\rm H_2}$, SFE$_{\rm ato}$ = SFR/M$_{\rm HI}$, SFE$_{\rm tot}$ = SFR/(M$_{\rm HI}$+M$_{\rm H_2}$). 
Similarly, we define the molecular, atomic and total gas fractions as: 
$f_{\rm gas,mol}$ = M$_{\rm H_2}$/(M$_{\rm H_2}$+M$_{\rm HI}$+M$_{\star}$), 
$f_{\rm gas,ato}$ = M$_{\rm HI}$/(M$_{\rm H_2}$+M$_{\rm HI}$+M$_{\star}$) and $f_{\rm gas,tot}$ = (M$_{\rm HI}$+M$_{\rm H_2}$)/(M$_{\rm HI}$+M$_{\rm H_2}$+M$_{\star}$).

\begin{figure*}
	\includegraphics[width=0.99\textwidth]{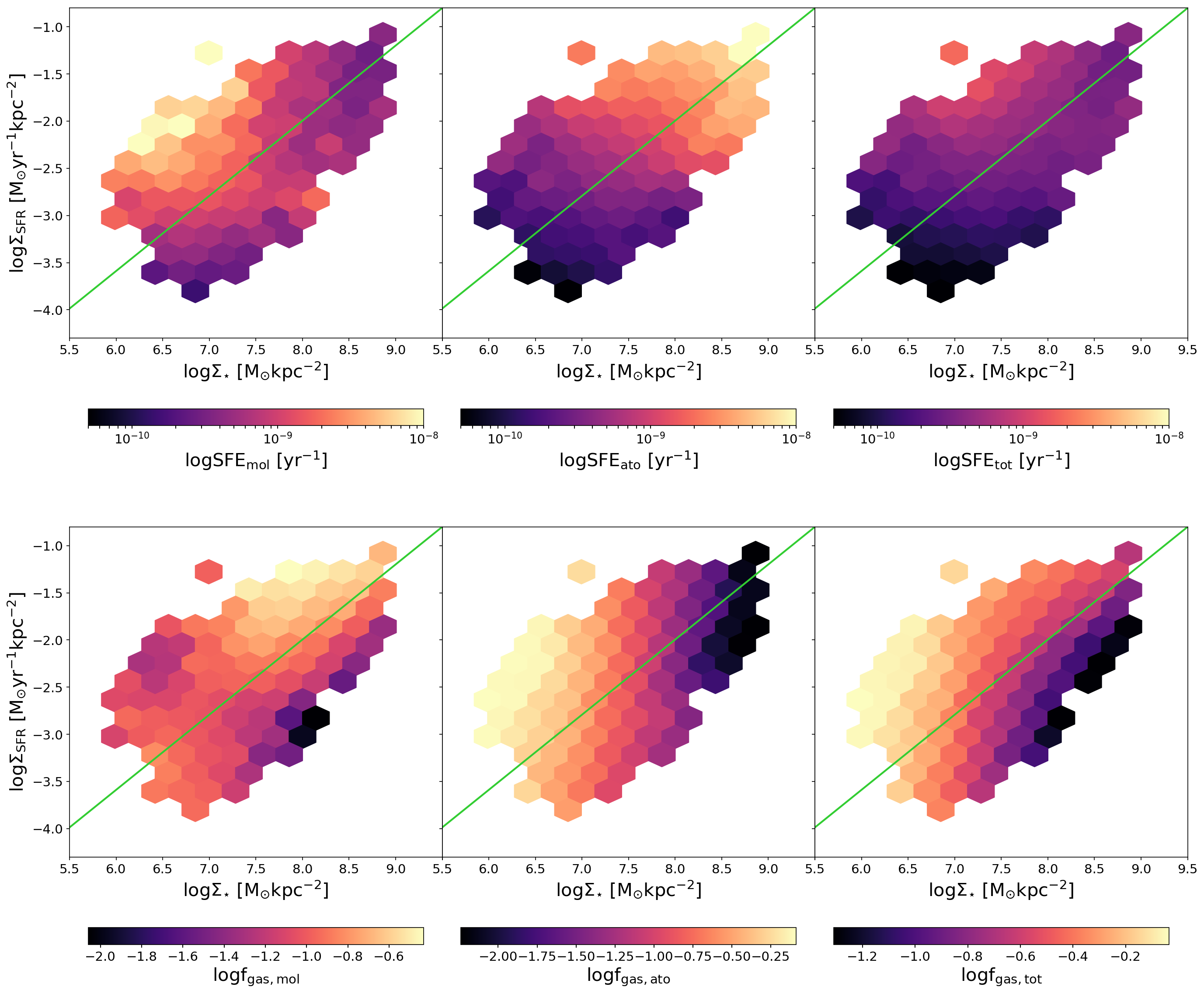}
    \caption{ Molecular, atomic and total SFE and $f_{\rm gas}$ in the log$\Sigma_{\star}$ - log$\Sigma_{\rm SFR}$ plane. The bins in the six panels are color coded as a function of: SFE$_{\rm mol}$ = SFR/M$_{\rm H_2}$ in the \textit{top left panel},  SFE$_{\rm ato}$ = SFR/M$_{\rm HI}$ in the \textit{top central panel},  SFE$_{\rm tot}$ = SFR/(M$_{\rm HI}$+M$_{\rm H_2}$) in the \textit{top right panel}, $f_{\rm gas,mol}$ = M$_{\rm H_2}$/(M$_{\rm HI}$+M$_{\rm H_2}$+M$_{\star}$) in the \textit{bottom left panel}, $f_{\rm gas,ato}$ = M$_{\rm HI}$/(M$_{\rm HI}$+M$_{\rm H_2}$+M$_{\star}$) in the \textit{bottom central panel}, and $f_{\rm gas,tot}$ = (M$_{\rm HI}$+M$_{\rm H_2}$)/(M$_{\rm HI}$+M$_{\rm H_2}$+M$_{\star}$) in the \textit{bottom right panel}. The spatially resolved MS is indicated with the green solid line.}
    \label{fig:sfe}
\end{figure*}

In Fig.\,\ref{fig:sfe} we show how the average SFE (top) and $f_{\rm gas}$ (bottom) vary in the log$\Sigma_{\star}$ - log$\Sigma_{\rm SFR}$ plane, separating the different phases. Qualitatively, we observe that SFE$_{\rm mol}$ varies strongly above and below the MS, for log$\Sigma_{\star}<$8M$_{\odot}$kpc$^{-2}$, while it is almost constant at higher stellar surface densities. These trends reflect the changes in $f_{\rm gas, mol}$, that is characterised by strong variations above/below the MS only at higher stellar surface densities, leading to a constant SFE. We notice that regions above the MS at log$\Sigma_{\star}<$7M$_{\odot}$kpc$^{-2}$ have very low $f_{\rm gas, mol}$, as they are located in the outer optical disc. SFE$_{\rm ato}$ (central top panel) is nearly constant above/below the MS, while it steadily increases along the MS relation for increasing stellar surface densities. The atomic gas fraction is subject to two distinct trends: it decreases along the MS relation,
from low to high $\Sigma_{\star}$, and tends to be higher above the MS. When considering the total gas, the variations of SFE$_{\rm tot}$ are less evident: SFE$_{\rm tot}$ increases slightly at fixed stellar surface densities, less significantly in the direction perpendicular to the MS relation, and it increases along the MS for log$\Sigma_{\star}<$7.5M$_{\odot}$kpc$^{-2}$, to reach an almost constant value for log$\Sigma_{\star}>$7.5M$_{\odot}$kpc$^{-2}$. Finally, we observe that $f_{\rm gas,tot}$ varies strongly when moving from the lower to the upper envelope of the MS, both perpendicular to the relation and at fixed log$\Sigma_{\star}$. Regions with log$\Sigma_{\star} < 7.0$ M$_{\odot}$, that on average correspond to log$\Sigma_{\rm H_2}$ below the sensitivity limit, have large gas fractions thanks to the contribution of the neutral gas.

\begin{figure*}
	\includegraphics[width=0.99\textwidth]{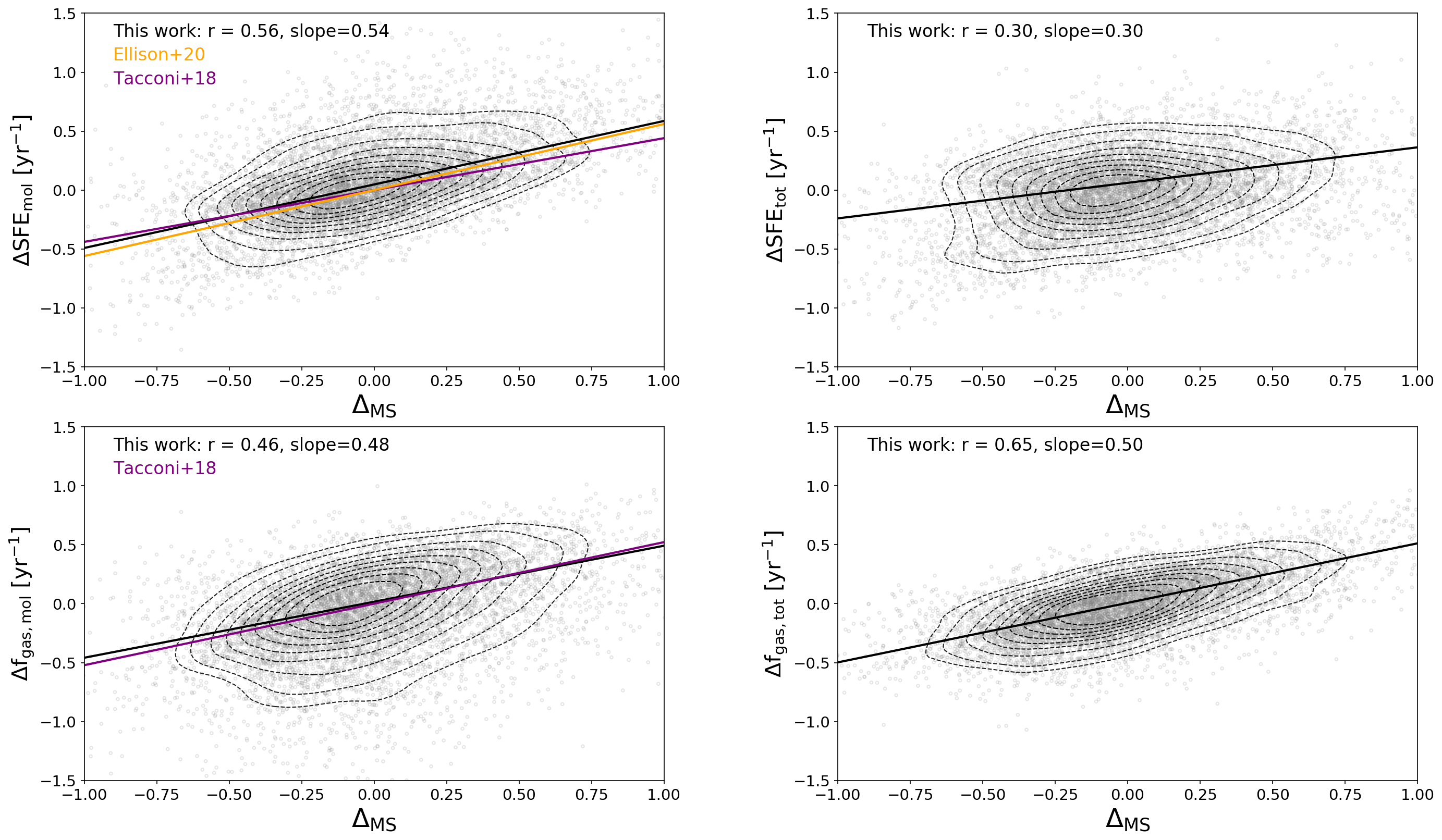}
    \caption{Distributions of regions in the $\Delta_{\rm SFE,mol}$ - $\Delta_{\rm MS}$ plane ({\it top left panel}), in the $\Delta_{\rm SFE,tot}$ - $\Delta_{\rm MS}$ plane ({\it top right panel}), $\Delta_{\rm f_{\rm gas,mol}}$ - $\Delta_{\rm MS}$ plane ({\it bottom left panel}), and $\Delta_{\rm f_{\rm gas,tot}}$ - $\Delta_{\rm MS}$ ({\it bottom right panel}). The best-fit relations of this work are shown as a black solid line. The best fit slope and the Pearson correlation coefficient are witting each panel. In the left panels we show relations of Tacconi et al. (2018) in purple, and the one of  Ellison et al. (2020) in orange. }    \label{fig:ellison}
\end{figure*}

The qualitative analysis carried on so far on Fig.\,\ref{fig:sfe} points to the role of both the SFE and $f_{\rm gas}$ in regulating the SFR of a region. From an integrated point of view, there is now convergence on the fact that an increase in SFR at fixed M$_{\star}$ and cosmic epoch is due to a combination of increasing gas mass and decreasing depletion time \citep[thus increasing SFE][]{2011MNRAS.415...61S, 2012ApJ...758...73S, 2013AJ....146...19L, 2014MNRAS.443.1329H}. In the recent years, these studies could be carried out with larger and larger samples, using different sub-mm observations to trace gas, and exploring a wide range of cosmic epochs, $0<z<4.5$, with the general consensus that an almost equal increase of gas fraction and SFE can explain an increase in sSFR at fixed redshift. \citet{2017ApJ...837..150S} exploited sub-mm ALMA continuum observations of $700$ COSMOS galaxies at $0.3<z<4$ and estimated the gas mass from the dust mass. Several works in literature indicate that the dust mass is a tracer of the total gas mass \citep[e.g.][]{2011ApJ...737...12L,2012A&A...542A..32C,2017A&A...602A..68O,2020A&A...633A.100C}, but for $0.3<z<4$ the total gas is dominated by the molecular fraction \citep[e.g.][]{2014MNRAS.443.1002L}. \citet{2015ApJ...800...20G} and \citet{2018ApJ...853..179T} combine three different estimates of molecular gas mass (from FIR SED, $\sim$1mm dust photometry and CO line flux) in the redshift range 0$<$z$<$4 and found that an increase in sSFR at fixed stellar mass and redshift is accompanied by an almost equal increase of SFE and $f_{\rm gas}$. Recently, \citet{2020MNRAS.493L..39E}, using molecular gas observations from the ALMaQUEST survey (median redshift $\sim0.03$) on kpc scales, find that variations of the SFE play a major role in setting the SFR at fixed stellar mass (and thus the scatter of the spatially resolved MS), with differences in $f_{\rm gas,mol}$ playing a secondary role.  In Fig.\,\ref{fig:ellison} we compare our results to the trends found in \citet{2018ApJ...853..179T} and \citet{2020MNRAS.493L..39E}. We define $\Delta_{\rm SFE}$ as the distance of a region from the spatially resolved KS law at fixed molecular gas mass (central panel of Fig. \ref{fig:bigiel}), and and $\Delta_{\rm f_{\rm gas}}$ as the distance of a region from the spatially resolved MGMS (central panel of Fig. \ref{fig:mgms}), as done in \citet{2020MNRAS.493L..39E}. As the analysis of the connection between the scatter of these relations and $\Delta_{\rm MS}$ in Sections \ref{sec:mass_gas} and \ref{sec:ks} revealed that MS regions are located along the KS and MGMS relations, the definitions of $\Delta_{\rm SFE}$ and  $\Delta_{\rm f_{\rm gas}}$  given above are consistent with those of \citet{2018ApJ...853..179T}, that normalise t$_{\rm depl}$ and $f_{\rm gas}$ to their MS values. The panels show the variation between molecular and total SFE and gas fraction of a region with respect to the MS values (thus $\Delta_{\rm SFE}$ and $\Delta_{\rm f_{\rm gas}}$) as a function of $\Delta_{\rm MS}$. The top left panel shows the $\Delta_{\rm SFE,mol}$ - $\Delta_{\rm MS}$ relation; the Pearson correlation coefficient is 0.56 and the best fit relation has a slope of 0.54 and a scatter of 0.18 dex, that is in excellent agreement with \citet[][slope = 0.44]{2018ApJ...853..179T} and \citet[][slope = 0.5]{2020MNRAS.493L..39E}. We emphasise here that our work has a spatial resolution of 500 pc, thus reaching surface densities that are $\sim$1 dex smaller than \citet{2020MNRAS.493L..39E}. A similar slope (0.48) is found when analysing $\Delta_{\rm f_{\rm gas,mol}}$ as a function of $\Delta_{\rm MS}$, with a Pearson coefficient of 0.47 and a scatter of 0.23 dex. Our best fit relation is in excellent agreement with the one of  \citet{2018ApJ...853..179T} (obtained from integrated quantities), that has a slope of 0.52. These results emphasise that when analysing the molecular gas phase, an increase in SFR at fixed M$_{\star}$ and cosmic epoch is due to a combination of increasing gas mass and increasing (decreasing) SFE (t$_{\rm depl}$). The situation painted by the molecular gas changes when the contribution of neutral gas is taken into account. In the left panels of Fig.\,\ref{fig:ellison} we show the $\Delta_{\rm SFE,tot}$ - $\Delta_{\rm MS}$ (top) and the $\Delta_{\rm f_{\rm gas,tot}}$ - $\Delta_{\rm MS}$ (bottom) planes, thus considering both the neutral and molecular gas phases. Interestingly, we see that the relation between $\Delta_{\rm SFE,tot}$ and $\Delta_{\rm MS}$ is significantly weaker than in the molecular case, as we retrieve a Pearson coefficient of 0.30, a slope of 0.30, and a scatter of 0.28 dex. On the other hand, we observe a stronger correlation between $\Delta_{\rm MS}$ and $\Delta_{\rm f_{\rm gas,tot}}$ (Spearman ranking is 0.65), with a slope of 0.50 and a scatter of 0.19 dex. These results indicate the importance of the neutral gas phase in setting the scatter of the spatially resolved MS, as it partially traces molecular gas dissociated by the radiation field. 

Finally, the dataset in our hands allow us to analyse the spatial variations of SFE and $f_{\rm gas}$ within the 5 galaxies of the sample. The spatially resolved maps of $\Delta_{\rm MS}$, $\Delta_{\rm SFE,tot}$ and $\Delta_{\rm f_{\rm gas,tot}}$ for every galaxy can be found in Appendix \ref{sec:a0} (Figures \ref{fig:distributions1} and \ref{fig:distributions2}). Here we emphasise that while we do observe variations within galaxies and among them of the connection between $\Delta_{\rm MS}$ and $f_{\rm gas}$ or SFE, on average the galaxy by galaxy analysis confirms that the gas fraction strongly correlates with $\Delta_{\rm MS}$.

\section{Conclusions}\label{sec:Conclusion}

In this manuscript we exploit the combination of highly accurate measurements of $\Sigma_{\star}$ and $\Sigma_{\rm SFR}$ at 500pc resolution of five nearby, face-on spiral galaxies obtained following the procedure presented in Paper I, with observations of neutral and molecular gas. With this powerful data set we study how the location of a region with respect to the spatially resolved {Main Sequence (MS)} is related to the gas in the different phases. We summarise here our main results: 

\begin{itemize}
\item We find that $\log\Sigma_{\star}$, $\log\Sigma_{\rm SFR}$, and $\log\Sigma_{\rm {H_2}}$ define a 3D relation (left panel of Fig.\,\ref{fig:3d}); the three projections are the Kennicutt-Schmidt (KS) law, the MS and the molecular gas main sequence (MGMS). The KS law is the tightest relation, with a scatter of 0.19 dex, followed by the MGMS (0.22 dex) and the spatially resolved MS (0.23 dex). The existence of the MGMS at sub-kpc scales opens up the possibility to study molecular gas content from log$\Sigma_{\rm SFR}$ and log$\Sigma_{\star}$ alone, that are generally easier to obtain and available for large samples;

\item We study the distribution of the neutral and molecular gas in the log$\Sigma_{\star}$ - log$\Sigma_{\rm SFR}$ plane (Fig.\,\ref{fig:fig4}) and we find that the surface density of molecular gas steadily increases along the MS relation, but is almost constant perpendicular to it. The surface density of neutral gas, instead, is almost constant along the MS, and increases (decreases) in its upper (lower) envelope. On average, regions located in the upper envelope of the spatially resolved MS have $\Sigma_{\rm {HI}}\geq$10M$_{\odot}$pc$^{-2}$, that is the typical value for the HI to H$_2$ transition. The three variables $\log\Sigma_{\star}$, $\log\Sigma_{\rm SFR}$, and $\log\Sigma_{\rm {HI}}$ are distributed along the plane $\log\Sigma_{\rm{SFR}}$ = 0.97$\log\Sigma_{\star}$ + 1.99$\log\Sigma_{\rm{HI}}$ - 11.11 that has a scatter of 0.14 dex (right panel of Fig.\,\ref{fig:3d}); 

\item When moving towards high $\Sigma_{\rm SFR}$ at fixed stellar surface densities, the molecular gas fraction and molecular SFE both increase. On the other hand, when we consider the total gas, thus also the contribution of the neutral component, we observe a steep increase of the gas fraction towards high SFRs, accompanied by a weak increase of the total SFE (Fig.\,\ref{fig:ellison}).
\end{itemize}

Our results illustrate the intricate interplay between neutral and molecular gas, as it changes radially as a function of the distance from the center of galaxies, as well as locally depending on the sSFR. We argue that molecular gas dissociation plays an important role in setting the observed trends of neutral and molecular gas surface densities around the MS relation (Fig.\,\ref{fig:alvio}). Thus, high total gas surface densities favor the formation of molecular hydrogen clouds, which in turn promote star formation whose resulting UV radiation has the effect of dissociating the molecules, in a local baryon cycle that is a manifestation of the self-regulating nature of the star formation process. We shall return on these issues in a future paper, also expanding on their implications for our understanding of the star-formation process in high-redshift galaxies. These trends, obtained here for massive, disc-dominated spirals without a strong bar, will be analysed for galaxies with different morphologies (bulge dominated and spirals with strong bars) in another upcoming Paper.

The continuity of trends above the MS, MGMS, and KS relations suggests that starburst regions do not result from a bi-modality in the star formation process, but rather from a steady variation of primarily the total gas fraction and partially the star formation efficiency. We speculate that this continuity could also explain the existence of high redshift starbursts, as the scatter of the spatially resolved MS is similar to the one of the integrated relation, and this last quantity is observed to be constant with redshift. It is widely believed that high redshift galaxies are dominated by H$_2$ over HI, because their higher gas surface density favors the molecular phase. However, this finding suggests that the higher sSFR at high redshift may actually contrast this expectation, with intense stellar feedback leading to dissociation thus reducing the H$_2$/HI ratio. The next generation of radio telescopes like the Square Kilometer Array ({\it SKA}), but also ongoing surveys with MeerKAT, such as MIGHTEE \citep{2016mks..confE...6J} and LADUMA \citep{2016mks..confE...4B}, will directly measure the HI content in distant galaxies, thus unraveling its role in the star formation processes at earlier cosmic epochs.

\section*{Acknowledgements}
We thank the anonymous referee for constructive comments that improved the manuscript. LM is grateful to Sara Ellison and Bhaskar Agarwal for helpful discussion on the manuscript. LM acknowledges support from the BIRD 2018 research grant from the Universit\'a degli Studi di Padova. AE and GR are supported from the STARS@UniPD grant. GR acknowledges the support from grant PRIN MIUR 2017 - 20173ML3WW\_001. GR and CM acknowledge funding from the INAF PRIN-SKA 2017 programme 1.05.01.88.04. We acknowledge funding from the INAF main stream 2018 programme "Gas-DustPedia: A definitive view of the ISM in the Local Universe". This research made use of {\sc photutils}, an Astropy package for detection and photometry of astronomical sources \citep{Bradley_2019_2533376}.

\section*{Data Availability}

 This research is based on observations made with the Galaxy Evolution Explorer, obtained from the MAST data archive at the Space Telescope Science Institute, which is operated by the Association of Universities for Research in Astronomy, Inc., under NASA contract NASA 5-26555. This work made use of HERACLES, ’The HERACO-Line Extragalactic Survey’ \citep{2009AJ....137.4670L} and THINGS,’The HI Nearby Galaxy Survey’  \citep{2008AJ....136.2563W}. DustPedia is a collaborative focused research project supported by the European Union under the Seventh Framework Programme (2007-2013) call (proposal no. 606847). The participating institutions are: Cardiff University, UK; National Observatory of Athens, Greece; Ghent University, Belgium; Universit\'e Paris Sud, France; National Institute for Astrophysics, Italy and CEA (Paris), France. We acknowledge the usage of the HyperLeda database (http://leda.univ-lyon1.fr). The derived data underlying this article will be shared on reasonable request to the corresponding author.



\bibliography{main}{}
\bibliographystyle{mnras}


\appendix
\section{The $\log\Sigma_{\star}$ - $\log\Sigma_{\rm SFR}$ - $\log\Sigma_{\rm {H_2,HI}}$ 3D relation}
\label{sec:a1}

\begin{figure*}
	\includegraphics[width=0.45\textwidth]{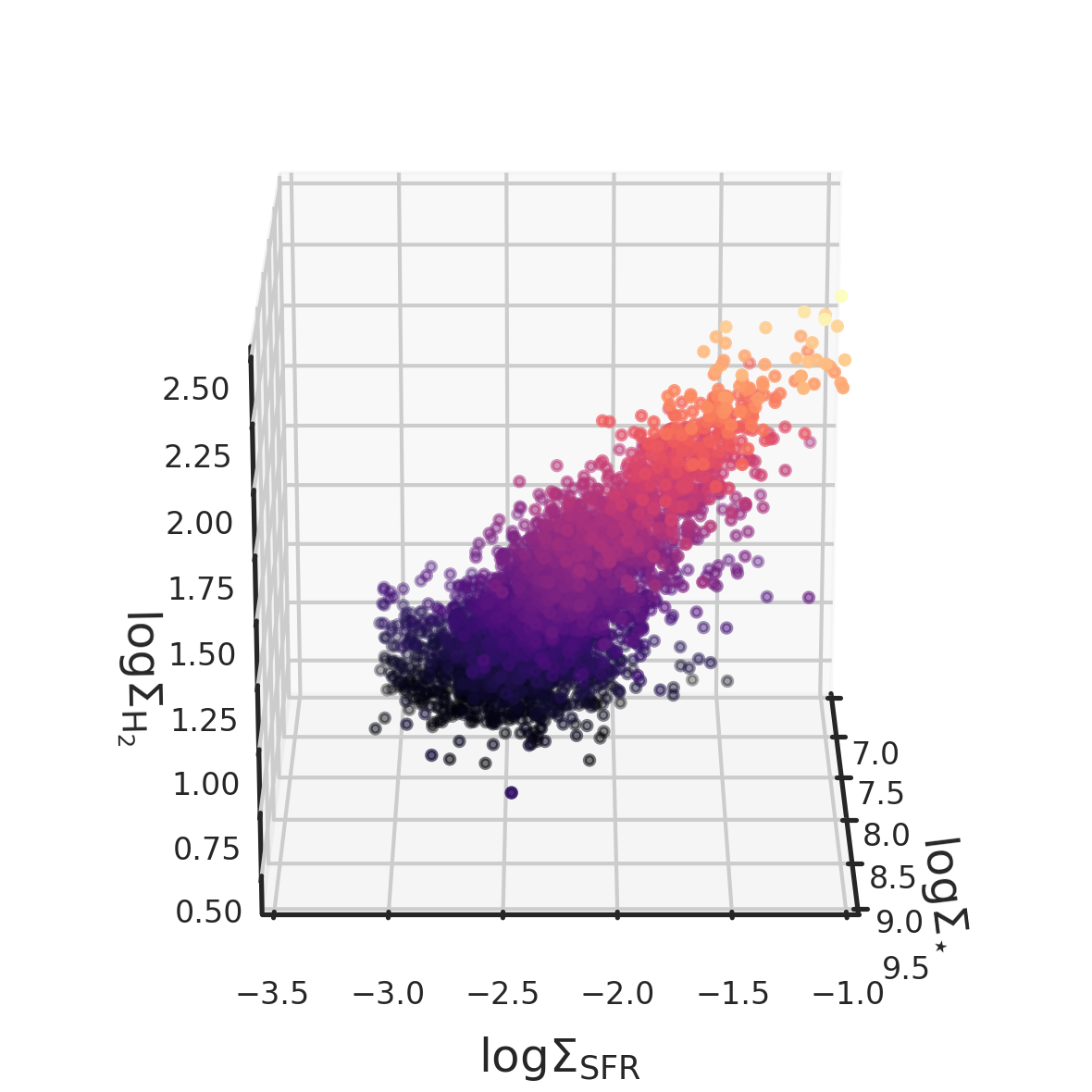}
		\includegraphics[width=0.45\textwidth]{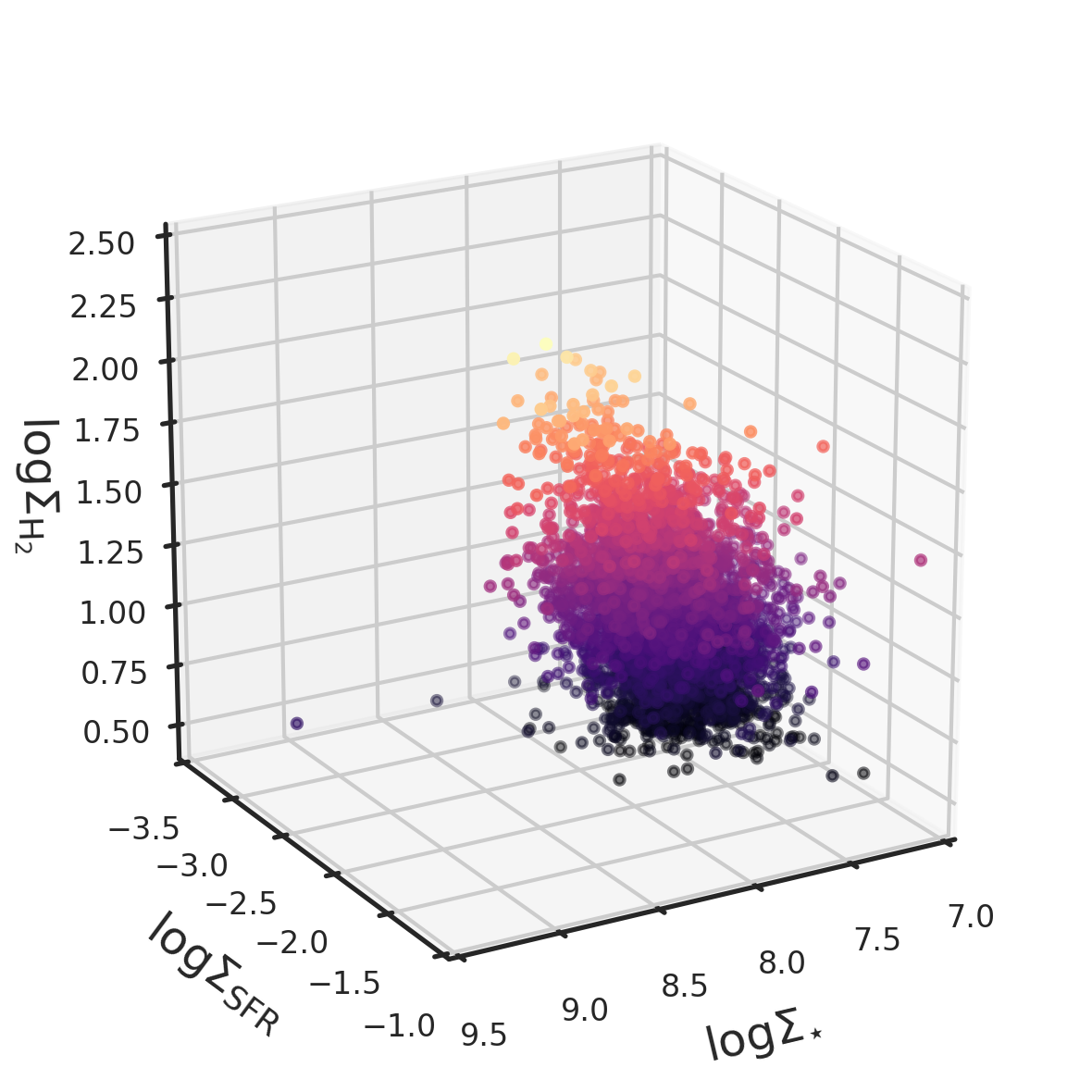}
			\includegraphics[width=0.45\textwidth]{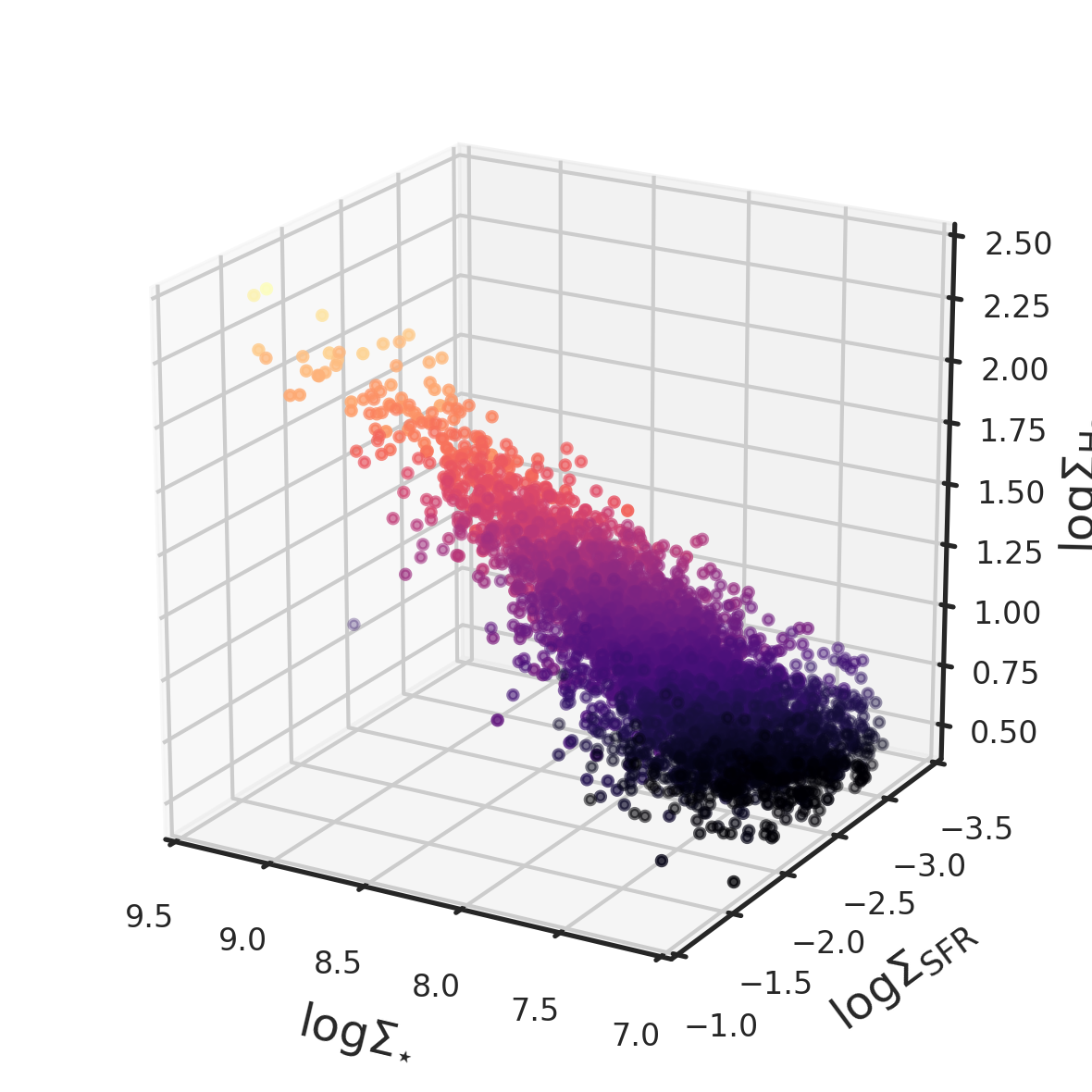}
				\includegraphics[width=0.45\textwidth]{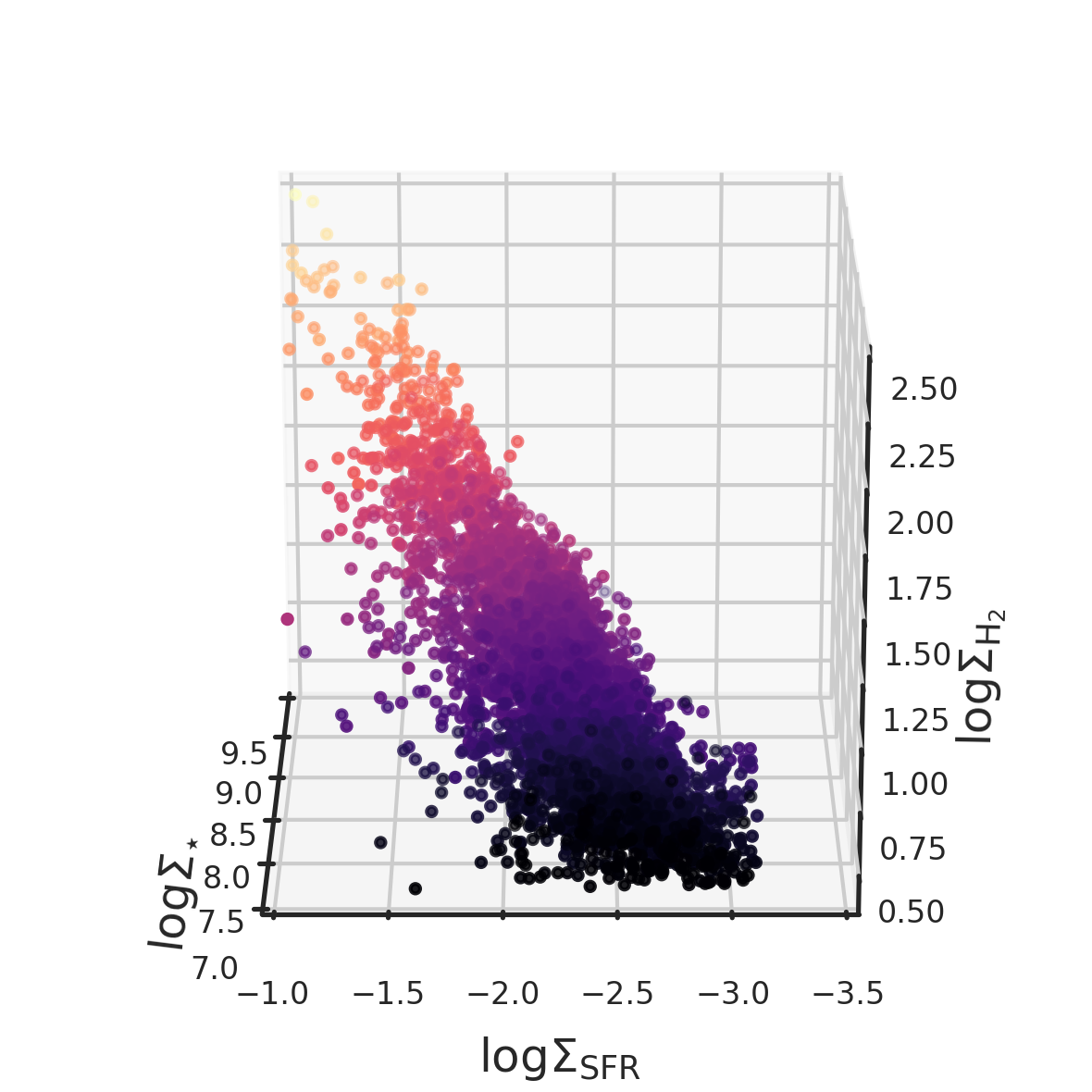}
			    	\includegraphics[width=0.6\textwidth]{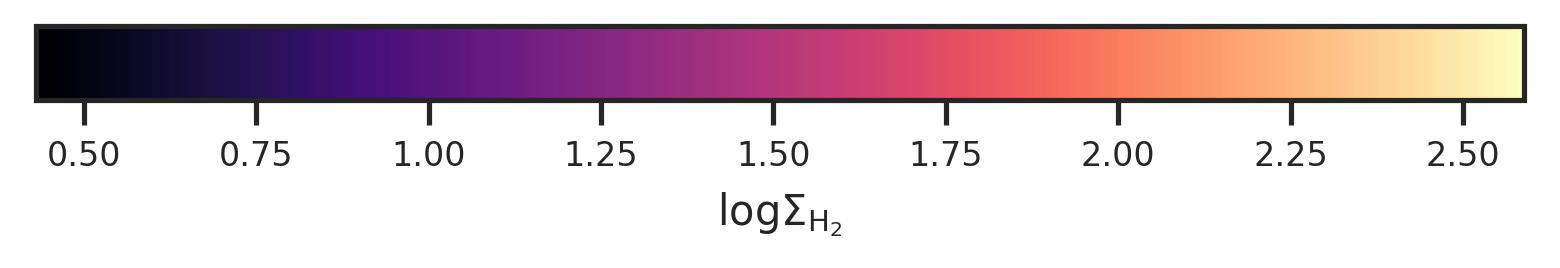}
    \caption{Distribution of the regions with an estimate of $\log\Sigma_{\rm {H_2}}$ above the sensitivity limit in the $\log\Sigma_{\star}$ - $\log\Sigma_{\rm SFR}$ - $\log\Sigma_{\rm {H_2}}$ 3D space. For different projections of the plane are shown, corresponding to different azimuthal angles: 0$^\circ$ (top left), 60$^\circ$ (top right), 120$^\circ$ (bottom left) and 180 $^\circ$ (bottom right). The point are colour coded as a function of $\log\Sigma_{\rm {HI}}$. } 
    \label{fig:sausage}
\end{figure*}

\begin{figure*}
	\includegraphics[width=0.48\textwidth]{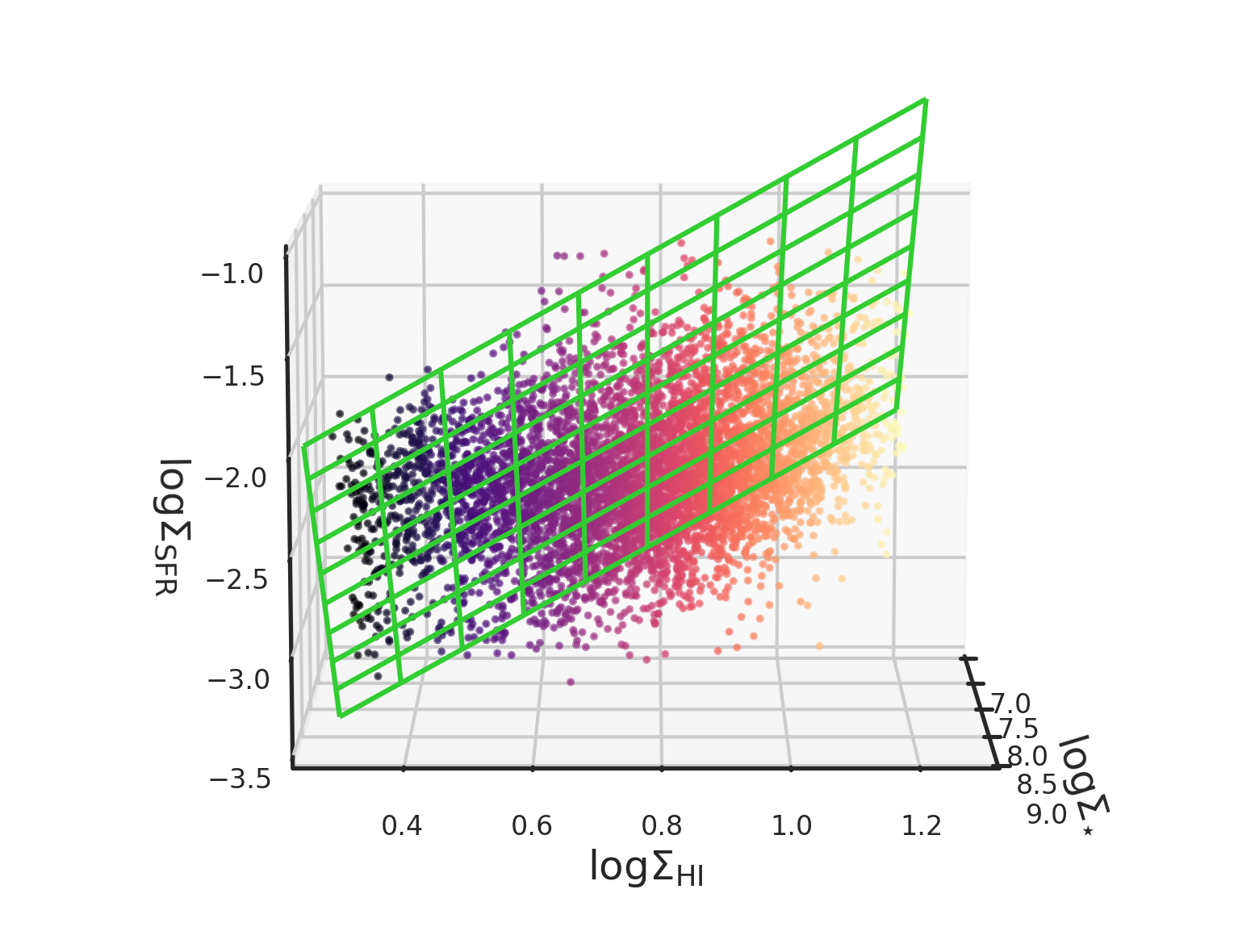}
		\includegraphics[width=0.48\textwidth]{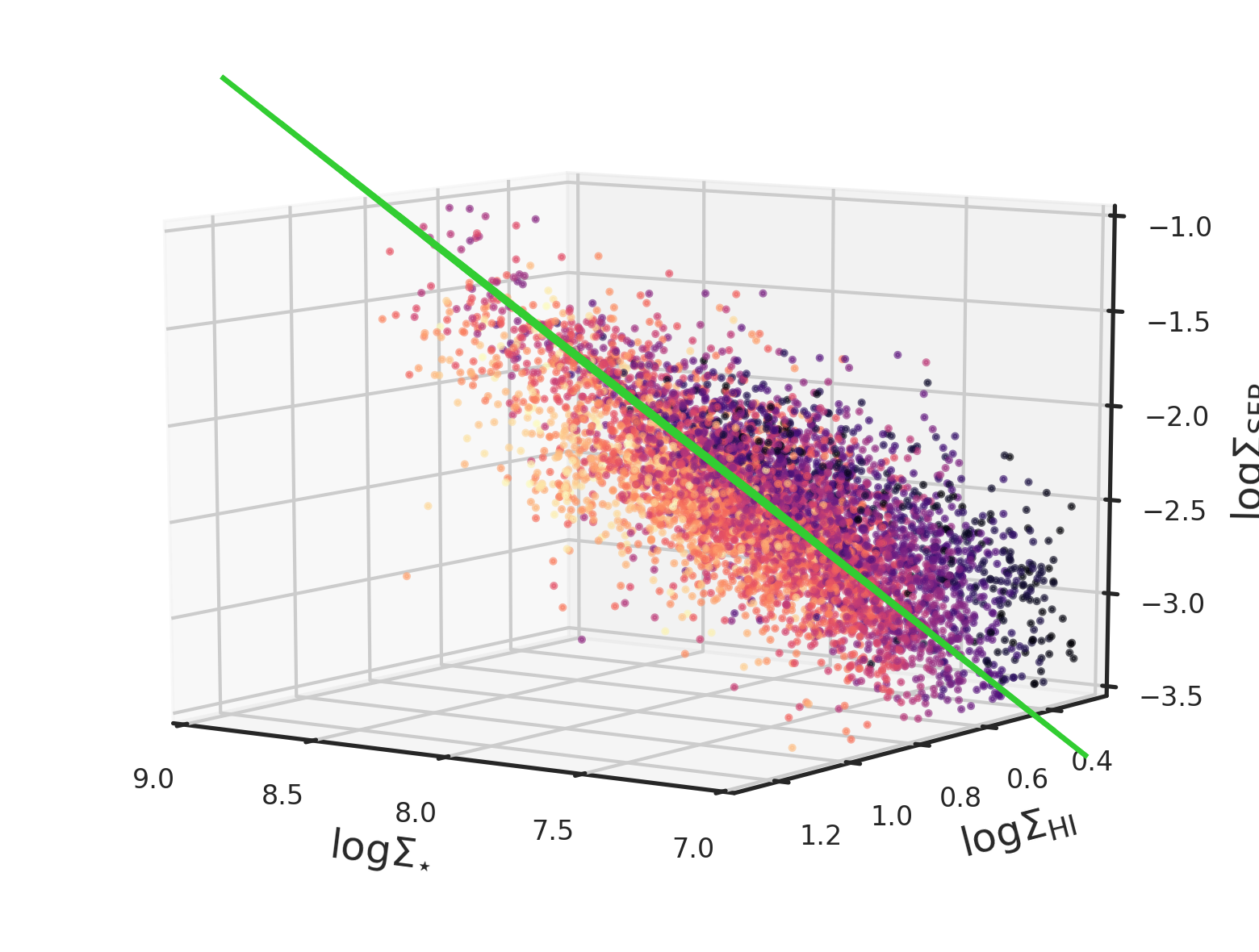}
			\includegraphics[width=0.48\textwidth]{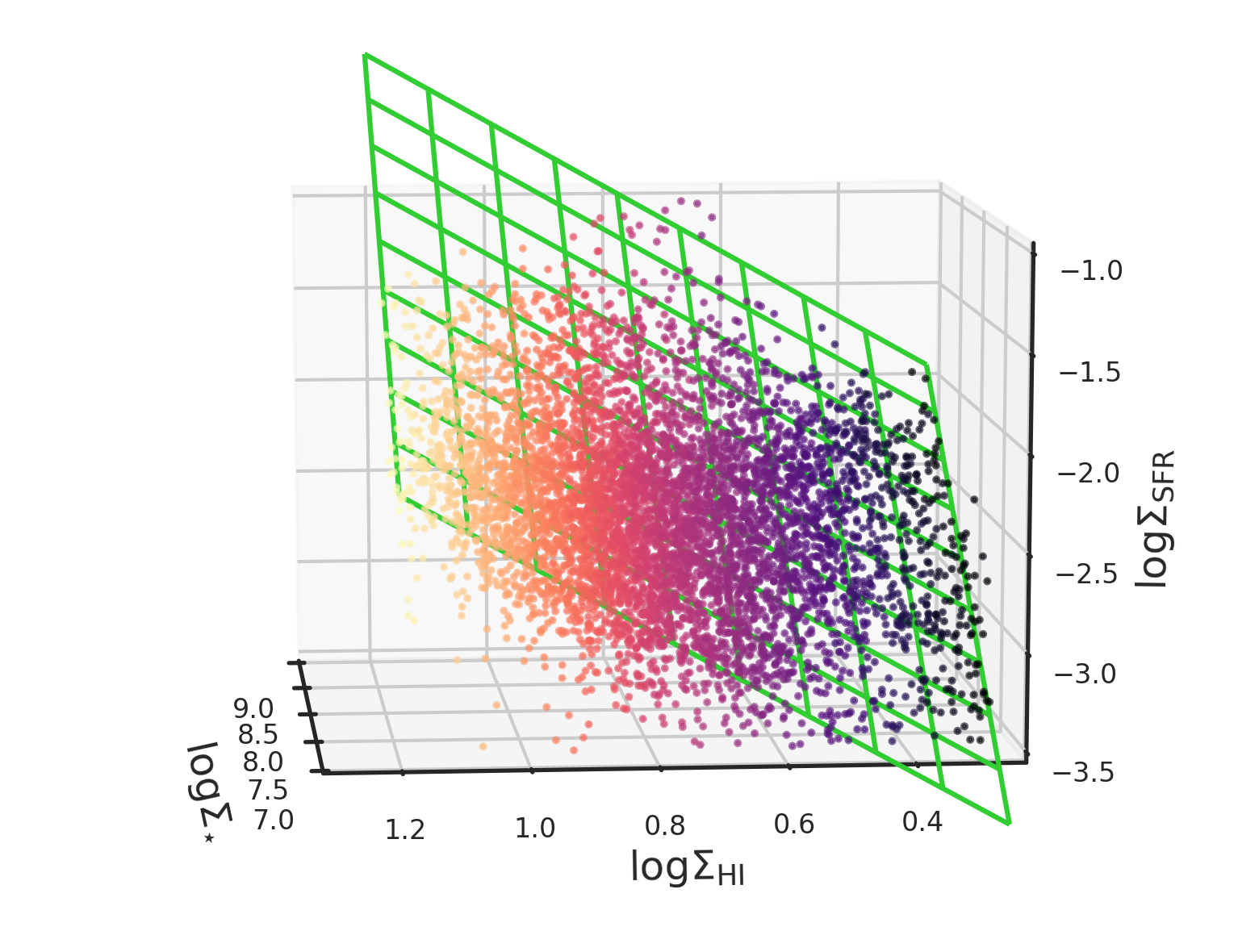}
				\includegraphics[width=0.48\textwidth]{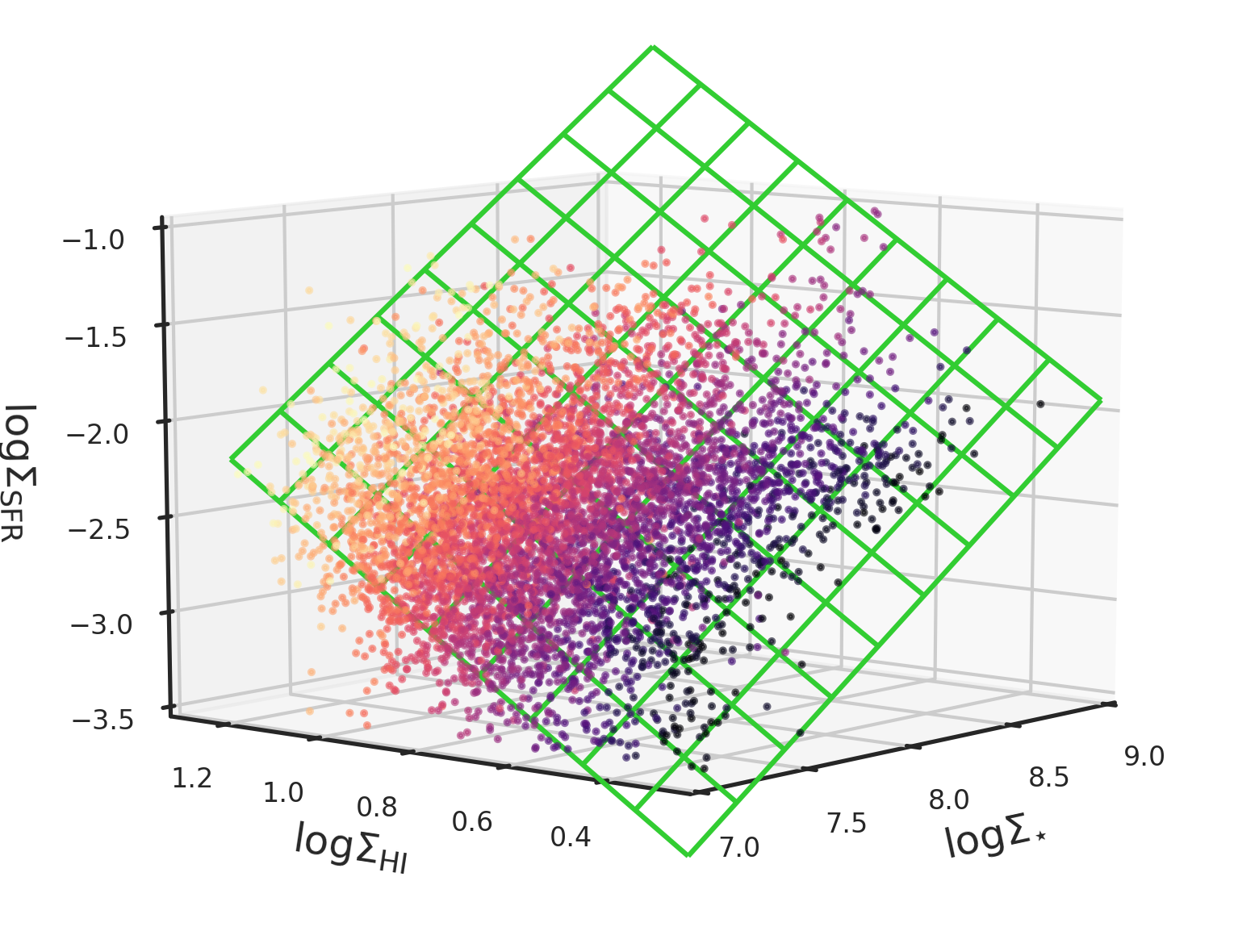}
			    	\includegraphics[width=0.6\textwidth]{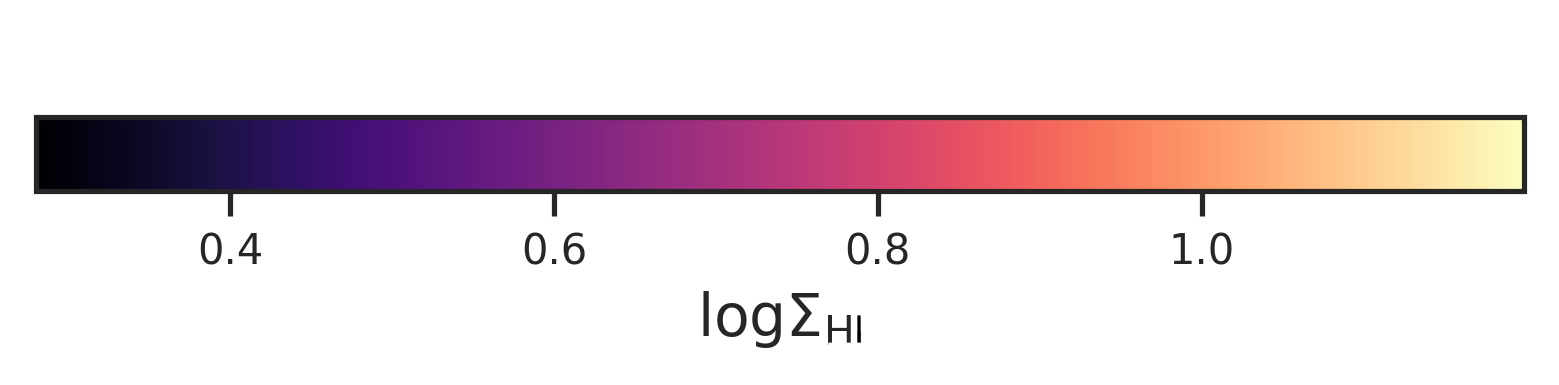}
    \caption{Distribution of the regions with an estimate of $\log\Sigma_{\rm {HI}}$ above the sensitivity limit in the $\log\Sigma_{\star}$ - $\log\Sigma_{\rm SFR}$ - $\log\Sigma_{\rm {H_I}}$ 3D space. For different projections of the plane are shown, corresponding to different azimuthal angles: 0$^\circ$ (top left), 125$^\circ$ (top right), 175$^\circ$ (bottom left) and 220 $^\circ$ (bottom right). The point are colour coded as a function of $\log\Sigma_{\rm {H_2}}$. } 
    \label{fig:plane}
\end{figure*}

As shown in Fig.\,\ref{fig:3d}, the three quantities $\log\Sigma_{\star}$, $\log\Sigma_{\rm SFR}$ and $\log\Sigma_{\rm {H_2}}$ define a 3D relation. To better visualise this relation, in Fig.\,\ref{fig:sausage} we show four different projections of it, corresponding to azimuthal angles of 0$^\circ$, 60$^\circ$, 120$^\circ$ and 180$^\circ$. We stress here that we are plotting only the regions that have an estimate of $\Sigma_{\rm {H_2}}$ above the sensitivity limit. In Fig.\,\ref{fig:plane} we show, instead, the $\log\Sigma_{\star}$, $\log\Sigma_{\rm SFR}$ and $\log\Sigma_{\rm {HI}}$ 3D plane at four different azimuthal angles, with the aim of better visualise the positioning of the cells along the 3D plane marked in green. We plot in Fig.\,\ref{fig:plane} only the regions that have an estimate of $\Sigma_{\rm {HI}}$ above the sensitivity limit.

\section{$\Delta_{\rm MS}$, SFE and gas content within galaxies}
\label{sec:a0}

\begin{figure*}
	\includegraphics[width=0.9\textwidth]{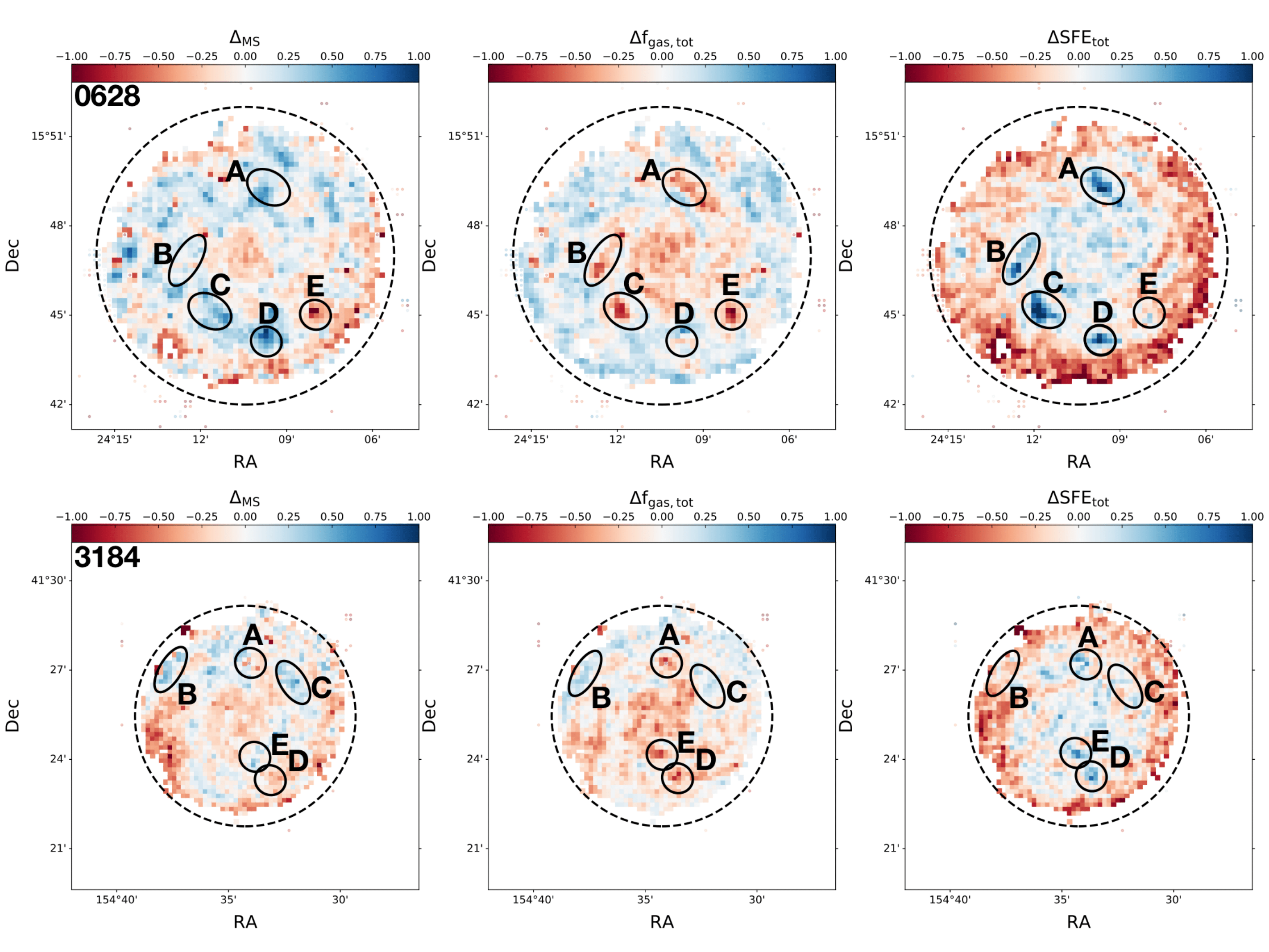}
			\includegraphics[trim = 0 0 0 0.7cm, width=0.9\textwidth]{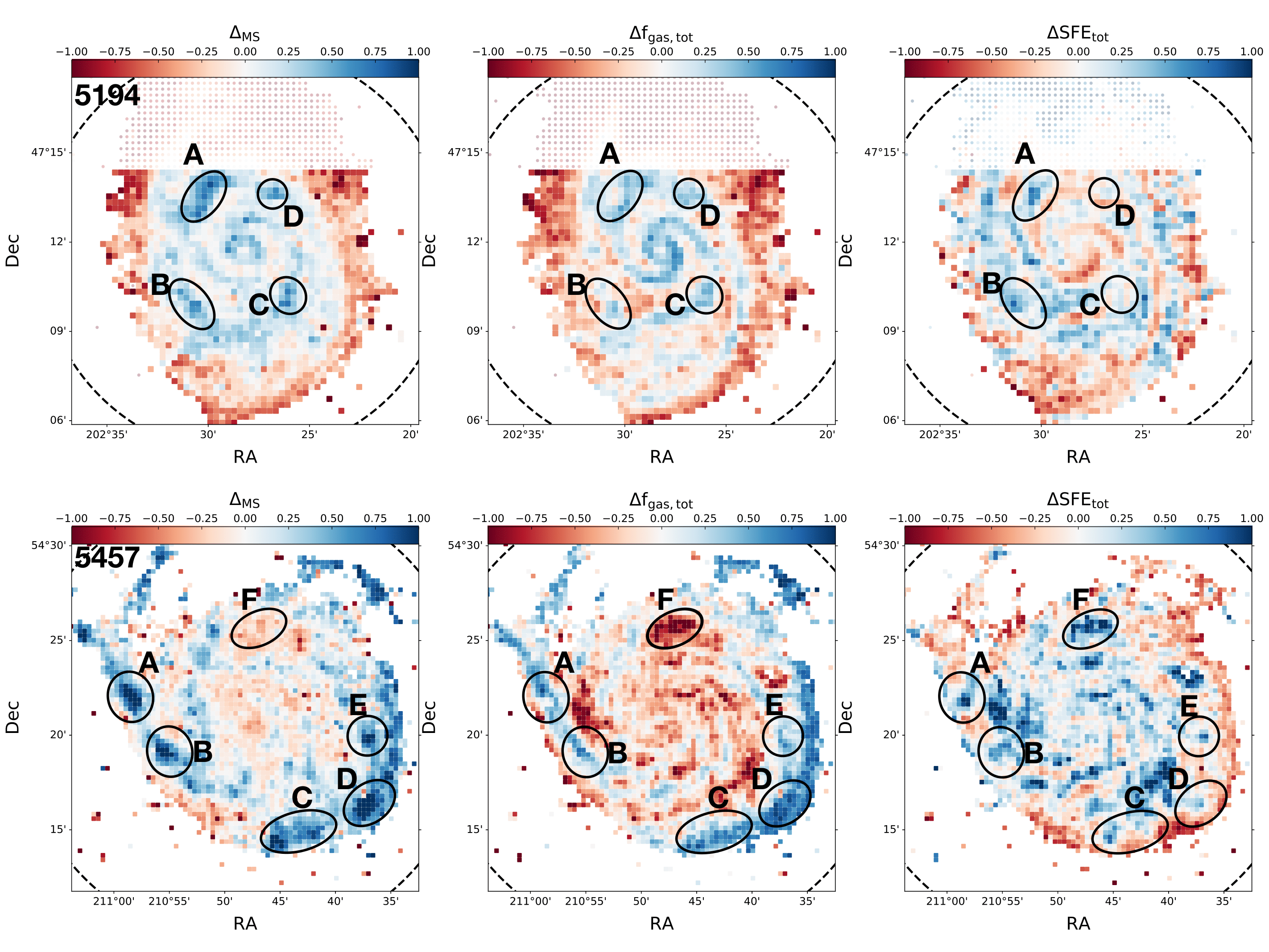}
    \caption{$\Delta_{\rm MS}$ (left), $\Delta_{\rm f_{\rm gas,tot}}$ (centre) and $\Delta_{\rm SFE,tot}$ (right) maps of (from top to bottom) NGC0628, NGC3184, NGC5194, and NGC5457. The highlighted regions are discussed more in detail in the text.}
    \label{fig:distributions1}
\end{figure*}

\begin{figure*}
	\includegraphics[width=0.92\textwidth]{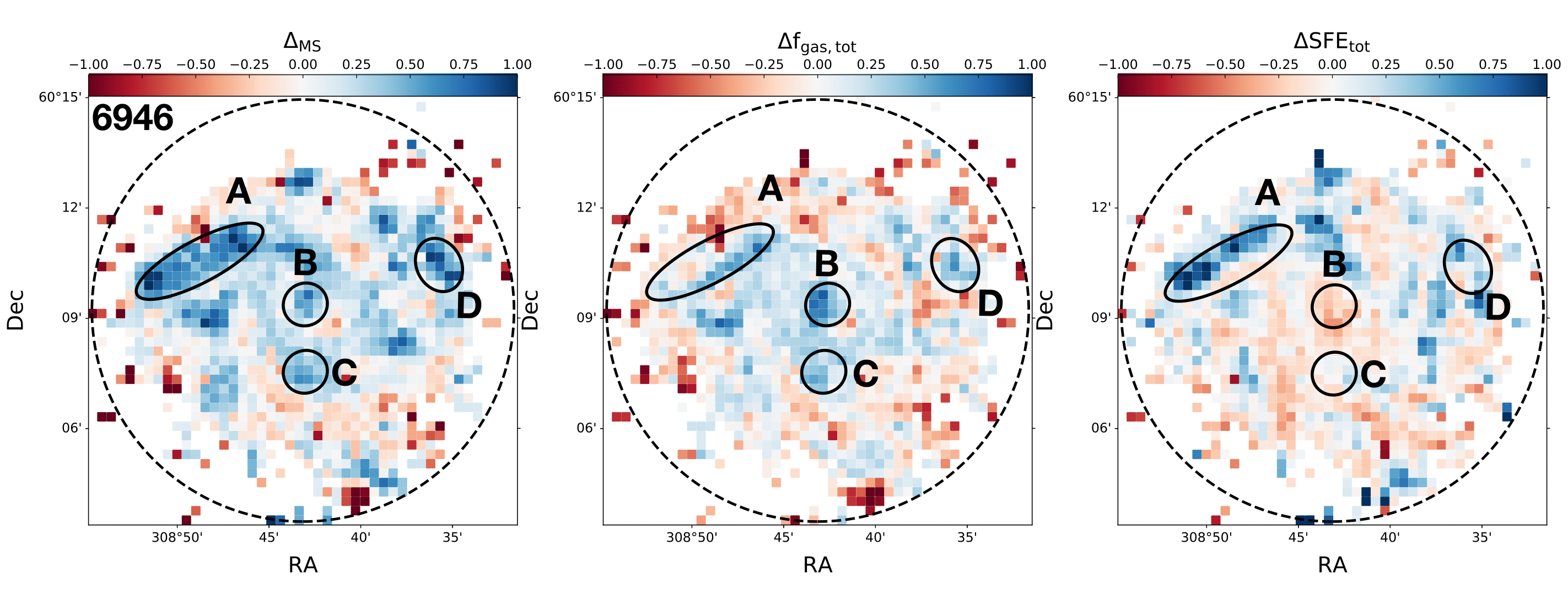}
    \caption{$\Delta_{\rm MS}$ (left), $\Delta_{\rm f_{\rm gas,tot}}$ (centre) and $\Delta_{\rm SFE,tot}$ (right) maps of NGC6946. The highlighted regions are discussed more in detail in the text.}
    \label{fig:distributions2}
\end{figure*}

In this Section, we discuss the distributions of $\Delta_{\rm MS}$, $\Delta_{\rm f_{\rm gas,tot}}$ and $\Delta_{\rm SFE,tot}$ within the five galaxies in our sample, shown in Figures\,\ref{fig:distributions1} and \ref{fig:distributions2}. We recall here that $\Delta_{\rm SFE,tot}$ and $\Delta_{\rm f_{\rm gas,tot}}$ are computed as the distance of the region from the global total gas - SFR relation (right panel of Fig. 2) and from the global total gas-mass relation (right panel of Fig. 3), and are thus a measure of the SFE and gas content with respect to the value on the best fitted relation at a given gas or stellar surface density (e.g. regions with $\Delta_{\rm f_{\rm gas,tot}}>$ 0 have higher gas content than the average). 

It is clear that the $\Delta_{\rm SFE,tot}$ and the $\Delta_{\rm f_{\rm gas,tot}}$ in the inner regions will be anti correlated because the regions where the bulk of molecular hydrogen is are not spatially coincident with the regions where the FUV star formation tracer peaks. This is due to the time delay between the formation of molecular clouds and the exposed phase of star formation \citep{2017A&A...601A.146C}. In the inner regions of spiral galaxies the spiral pattern rotates slower than the stars and hence molecular clouds form out of compressed gas along the inner part of the arm. These are the regions with higher gas fraction than the average. The young stars become visible after dispersing the original gas and, moving faster than the arm, they appear shifted with respect to the newly formed molecular complexes. These are regions with low gas content but numerous visible young stars, hence appear with a star formation efficiency above the average. For the outer regions, at or beyond corotation, the situation should reverse with respect to the arm but the anticorrelation between $\Delta_{\rm SFE,tot}$ and the $\Delta_{\rm f_{\rm gas,tot}}$ should still hold. Here the low surface density of molecules and the presence of external perturbations make the study more challenging: tidal perturbations, gas stripping and infall have a strong impact and this reflects on $\Delta_{\rm MS}$ which can vary substantially from one region to the next, being star forming regions more coarsely spaced in the outer disks. Below we describe a few selected regions in each galaxy more in detail:

\begin{itemize}
\item The central region of NGC0628, i.e. the bulge, is located below the MS and has low gas content, while $\Delta_{\rm SFE,tot}$ varies of $\pm$0.3dex around zero. Regions located in the upper envelope of the MS relation are located mostly along the spiral arms. We observe regions where $\Delta_{\rm MS}$ $\sim0$ (A,B and C) or $\Delta_{\rm MS}\sim$ 1 (D), or $\Delta_{\rm MS}\sim$ -1 (E), all with low gas fractions and high or average SFE.

\item NGC3184 also has a central region characterised by mostly negative $\Delta_{\rm MS}$ and $\Delta_{\rm f_{\rm gas,tot}}$, while $\Delta_{\rm SFE,tot}$ is, for the majority of cells, positive. In the spiral arms, we observe regions where the SFR is significantly higher than the MS value (B and C) that correspond to gas fractions above the average, and varying SFE. We identify three blobs (A, D \& E) with negative $\Delta_{\rm f_{\rm gas,tot}}$ and $\Delta_{\rm MS}$ varying slightly around the MS value (as in A \& E) or $\Delta_{\rm MS}<0$ (as in D).

\item NGC5194 is characterised by a central region that has SFR values mostly above the MS. Positive values of $\Delta_{\rm MS}$ are located mainly along the spiral arms, that are also well traced by an excess of total gas compared to the average. The regions A, B, C and D are all above the MS relation, but they show different properties: while the majority of pixels within C and D are have $\Delta_{\rm f_{\rm gas,tot}}>0$ and $\Delta_{\rm SFE,tot}\sim0$, A and B show clumps of $\Delta_{\rm SFE,tot}$ well above 0 and $\Delta_{\rm f_{\rm gas,tot}}$ varying between negative and positive values.

\item NGC5457 is characterised by two strong spiral arms with several blobs that have a SFR higher than the MS value (regions A, B, C, D, E). Within these regions we observe gas fractions that are almost always higher than the MS values, while $\Delta_{\rm SFE,tot}$  varies more strongly between positive and negative values. Region F, for which the majority of cells are well below the MS relation, has $\Delta_{\rm SFE,tot}\sim1$, but $\Delta_{\rm f_{\rm gas,tot}}\sim$-1.

\item NGC6946 has a central region characterised by SFR higher than MS values, high gas content and SFE lower than the average (region B). Regions A, C and D are also characterised by values of $\Delta_{\rm MS}$ close to 1, but different SFE. While regions A and D have close to normal gas content and SFE above the average, the more central regions B and C have a a high gas content but a low SFE.

\end{itemize}

In general we conclude that in the innermost and outer regions the decrease or enhancement of the local gas content follows closely that of the star formation rate along the MS. For the innermost region and the spiral arms the enhancement or decrease of the SFE is opposite to that of the local gas content. But there are exceptions to this behaviour. Given the possible temporal and spatial lag between molecular and FUV peaks \citep{2019Natur.569..519K} the relation between $\Delta_{\rm SFE,tot}$  and $\Delta_{\rm MS}$ for individual galaxies is more complex and requires dedicated analysis.



\bsp
\label{lastpage}

\end{document}